\documentclass[letterpaper,amsmath,amssymb,floatfix,aps,superscriptaddress]{revtex4}
\usepackage{geometry}                
\usepackage{graphicx}
\usepackage{amssymb}
\usepackage{epstopdf}
\usepackage{hyperref}
\usepackage{color}
\usepackage{pstricks}
\newcommand{\xx}{{\boldsymbol{x}}}

\newcommand{\pp}{{\boldsymbol{p}}}
\newcommand{\kk}{{\boldsymbol{k}}}

\newcommand{\rr}{{\boldsymbol{r}}}

\newcommand{\aal}{{\boldsymbol{\alpha}}}

\renewcommand{\AA}{{\boldsymbol{A}}}
\newcommand{\BB}{{\boldsymbol{B}}}

\newcommand{\ind}[1]{_\textrm{#1}}

\newcommand{\epsext}{\epsilon_\textrm{ext}}
\newcommand{\deltext}{\varepsilon_\textrm{ext}}

\newcommand{\mZ}{\mathcal{Z}}

\begin{document}

\title{Electrostatic interactions mediated by polarizable counterions: \\
weak and strong coupling limits}

\date{}  

\author{Vincent D\' emery}
\affiliation{Institut Jean Le Rond d'Alembert, CNRS and UPMC Universit\'e Paris 6, UMR 7190, F-75005 Paris, France, EU}
 
\author{David S. Dean}
\affiliation{Universit\' e de Bordeaux, Laboratoire Ondes et Mati\` ere d'Aquitaine (LOMA), UMR 5798 F-33400 Talence, France, EU}

\author{Rudolf Podgornik}
\affiliation{Department of Theoretical Physics, J. Stefan Institute and Department of Physics, Faculty of Mathematics and Physics, University of Ljubljana - SI-1000 Ljubljana, Slovenia, EU}

\begin{abstract}
We investigate the statistical mechanics of an inhomogeneous Coulomb fluid composed of charged particles with static polarizability. We derive the weak- and the strong-coupling approximations and evaluate the partition function in a planar dielectric slab geometry with charged boundaries.
We investigate the density profiles and the disjoining pressure for both approximations. 
Comparison to the case of non-polarizable counterions shows that polarizability brings important differences in the counterion density distribution as well as the counterion mediated electrostatic interactions between charged dielectric interfaces. 
\end{abstract}

\maketitle

\section{Introduction}

It has become clear by now that the standard Poisson-Boltzmann (PB) theory used to describe and understand the 
electrostatic interactions in colloidal systems has severe limitations and can sometimes give qualitatively unreliable 
if not outright wrong answers \cite{Naji2010}. There are several distinct reasons why the PB theory cannot describe some salient features of  highly charged Coulomb systems.

First and most notably, the PB theory is a mean-field theory, and thus completely misses the important effects of ionic correlations that have recently been the focus of much research in Coulomb fluids \cite{Kanduc2009}. The correlation effect, first observed in simulations \cite{Guldbrand1984}, exposes the limitations of the mean-field \emph{ansatz} in quite a drastic manner, since for highly charged systems the interactions mediated by mobile ions between equally charged interfaces can become attractive. However, general theorems demand the interaction to be repulsive at the mean-field level \cite{Neu1999,Sader1999,Trizac1999}. Several lines of thought were spawned by simulations and converged into a paradigm shift that allowed for a simple conceptual understanding of why the mean-field picture breaks down for highly charged systems and how to formulate a theory that would be valid in these circumstances \cite{Naji2005}. This paradigm shift led to a dichotomy between the weak and strong-coupling approaches that delimit the exact behavior of a Coulomb system at any value of electrostatic coupling \cite{Kanduc2010}. 

Another drawback of the PB theory is the physical model on which it is based -- point charged particles -- that neglects all ion-specific effects except for the ion valency. It is thus a \emph{one parameter theory} where the ions differ only in the amount of charge they bear. One straightforward way to amend this drawback, sharing some of the conceptual simplicity with the original Poisson-Boltzmann formulation, is to take into account the excess static ionic polarizability of the ions \cite{Ben-Yaakov2011,Ben-Yaakov2011b,Frydel2011} proportional to the volume of the cavity created by the ion in the solvent. Static excess ionic polarizability is then a second parameter that differentiates between different but equally charged ionic species and thus presents an important step towards more {\em civilized} models of Coulomb fluids.  

Studies of the excess ionic polarizability have a venerable history and go all the way back to the classical book by Debye on polar molecules  (see discussion on pages 111-115 in Ref. \cite{Debye1929}), where he already discussed cavities around ions having a different value of dielectric constant compared to the surrounding solution. These cavities in fact represent excess polarization of the ions in aqueous solvent.  Since due to saturation effects for most salts the interior dielectric constant should be taken much smaller than the aqueous one, the corresponding (static) dielectric constant of the salt  should then be smaller than for pure solvent. This corresponds to negative excess ionic polarizability. While in Debye's analysis the effect scales as the volume of the ionic cavity, there are indications that for large enough solutes it should actually scale with the area of the cavity \cite{Chandler2005}.

One of the moot points of Debye's analysis is exactly how to pick the right size of the cavity, an issue that has continued unabashed ever since \cite{Conway1981}.  The changes in the effective dielectric constant of ionic solutions due to ionic polarizability was later picked up by  Bikerman \cite{Bikerman1942} who, among other things, acknowledged that a realistic treatment of ions in aqueous solution should take their finite size into account (see also the discussion in \cite{Hatlo2012})  as well as their excess polarizability.  The effects of ionic polarizability and the associated dielectric decrement on the interactions between charged macromolecular surfaces in the presence of mobile counterions has been investigated in more recent times starting from the fundamental work of Netz \cite{Netz2001} and continuing with a steady stream of works \cite{Ben-Yaakov2011, Ben-Yaakov2011b,Frydel2011,Hatlo2012}.

Some facets of the ionic and colloid polarizability were discussed starting from the weak-coupling level by generalising the zero Matsubara frequency van der Waals term and modifying the appropriately formulated linearised Debye-H\"uckel theory \cite{Netz2001,Netz2007}.  Levin and coworkers \cite{Levin2009a,Levin2009} dealt with polarizability in the context of (ideally) polarizable ions in the vicinity of the dielectric interface. They also formulated a theory of monovalent and multivalent counterions in suspensions of polarizable colloids or nanoparticles  \cite{Bakhshandeh2011}\ which in some respects complements our work where the mono or polyvalent counterions themselves are treated as polarizable.

The main conceptual fulcrum of our present work is the dielectric decrement of ionic solutions that has been attributed to various sources, which underlie the changes in the dielectric response of the solution, but can be universally quantified by an excess ionic polarizability  \cite{Ben-Yaakov2011,Ben-Yaakov2011b}. It is proportional to the derivative of the (static) dielectric constant of a salt solution with respect to the concentration of the ions. Numerically this last coefficient, $\tilde\beta$ \cite{Ben-Yaakov2011}, turns out to be between $-7\,{\rm M}^{-1}$   and $-20\,{\rm M}^{-1}$ for most of the common salts \cite{Hasted1973}. 
  
Here we shall proceed with the analysis of effects of the excess static ionic polarizability of ions by formulating  consistent weak- and strong-coupling approaches that will lead to a two parameter -- charge and static excess polarizability -- theory  of a Coulomb fluid. We thus reformulate the basic model of a Coulomb fluid and investigate its consequences. This is accomplished by first  incorporating the excess ionic polarizability effect in a consistent way into a microscopic model and then solving the corresponding theory at the mean-field weak-coupling level as well as at the strong coupling level. 
It further turns out that the radius of the ions (more precisely of their hydration shell or cavity) must be introduced, leading to a more civilized three parameters theory. The presented theory has thus a very broad parameter space that we can not analyze in complete detail. We point to some salient features and leave most of the details for future endeavor.

\section{Model}

\begin{figure}
 \begin{center}
\includegraphics[angle=0,width=0.5\linewidth]{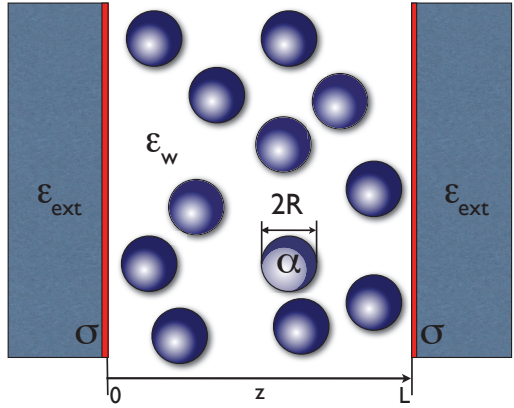}
\end{center}
\caption{Polarizable counter-ions of excess polarizability  $\alpha$ between two charged plates. The solvent in between has a permittivity $\epsilon_w$, while the two semi-infinite regions $0 > z > L$ have permittivity $\epsext$. The two surfaces at $z = 0, L$ bear a surface charge of surface charge density $\sigma$. $R$ is the radius of the ions.}
\label{f_system}
\end{figure}

We are interested in the behavior of mobile charges (counterions) 
immersed in a planar slab of thickness $L$ filled by aqueous solvent of permittivity $\epsilon\ind{w}$. The slab is assumed to be confined between two semi-infinite regions of permittivity $\epsext$ that bear fixed charges of opposite sign to the sign of the mobile charges with surface charge density $\sigma_0$. Counterions have a radius $R$, a charge $e = q e_0$, where $e_0$ is the elementary charge of the electron and $q$ is their valency,  and an excess polarizability $\alpha$. This latter quantity is defined precisely as the difference between the aqueous solvent polarizability and the proper ionic polarizability, and may thus be negative as surmised by Debye \cite{Debye1929}. In fact experimentally this is the standard behavior observed for many salts, see Ref \cite{Ben-Yaakov2011} for details. We will denote the whole space as $E$ and the volume of the slab as $V$. A schematic representation of the geometry of our model is given in Fig. \ref{f_system}. 

\subsection{Field-action}\label{}

The partition function for $N$ counterions is
\begin{eqnarray}\label{zn}
Z_N & = & \frac{1}{N!}\int[d\phi]\prod_{j=1}^N d\xx_j \\ & &\times \exp\left(-\frac{\beta\epsilon_0}{2}\int_E\epsilon(\xx)(\nabla\phi(\xx))^2 d\xx+\beta\sum_j\left[ie\phi(\xx_j)-\frac{\alpha}{2}(\nabla\phi(\xx_j))^2\right]-i\beta\int_{\partial V}\sigma(\xx)\phi(\xx)d'\xx\right),\nonumber 
\end{eqnarray}

where $\beta = (k_BT)^{-1}$ and $d'\xx$ denotes the integration over the bounding surfaces $\partial V$. The standard field-theoretical representation of the Coulomb fluid partition function in terms of the fluctuating electrostatic potential has been used \cite{Dean2009b}, properly extended by the fact that the counterion energy in an electrostatic field contains the point charge contribution $ie\phi(\xx_j)$ as well as the term due to its excess polarizability $\frac{\alpha}{2}(\nabla\phi(\xx_j))^2$.

Note that in this general expression, the surface charge may not be uniform, although we will restrict ourselves to the case $\sigma(\xx)=\sigma_0$.

The grand canonical partition function for a given fugacity $\lambda$ is then given by
\begin{equation} \label{xi}
\mZ=\sum_{N=0}^\infty \lambda^N Z_N= \int \exp(-\beta S[\phi])[d\phi],
\end{equation}
where the field-action $S[\phi]$ is given by

\begin{equation}
\beta S[\phi]=\frac{\beta\epsilon_0}{2}\int_E\epsilon(\xx)(\nabla\phi(\xx))^2 d\xx - \lambda\int_V \exp\left(-\beta\left[\frac{\alpha}{2}(\nabla\phi(\xx))^2-ie\phi(\xx)\right]\right) d\xx - i\beta\int_{\partial V} \sigma(\xx)\phi(\xx)d'\xx.
\label{xi}
\end{equation}
This is the fundamental expression that we will evaluate; we will specifically concentrate on its dependence on the separation between charged plane-parallel boundaries.

\subsection{Dimensionless field-action}\label{}

The field-action can be rewritten in terms of dimensionless parameters. Of course this analysis holds only in 3D; in other dimensions, the characteristic lengths that define the dimensionless parameters would have to be defined differently \cite{Dean2009b}. The dimensionless form of the action itself suggests various approximations that allow an explicit and exact evaluation of the grand canonical partition function \cite{Naji2005}.

First, we recall the definition of the Bjerrum and Gouy-Chapman lengths, $l\ind{B} =  {\beta e_0^2}/{4\pi \epsilon\ind{w}\epsilon_0} $ and $ l\ind{GC}  = {1}/ {2\pi q l\ind{B}\sigma_S}$, where $\sigma_0 =  e_0 \sigma_S$ is chosen to be positive. The electrostatic "coupling constant" is then defined as the ratio \cite{Boroudjerdi2005}
\begin{equation}
\Xi = \frac{q^2 l\ind{B}}{l\ind{GC}} = 2\pi q^3 l\ind{B}^2\sigma_S = q^3 ~\Xi_0.
\end{equation}
Above we specifically decomposed the coupling parameter into its $q$ and $\sigma_S$ dependence. The dimensionless length, field, permittivity and surface charge can then be expressed as  $\tilde \xx  =  \xx/l\ind{GC}$, $ \tilde\phi  =  \beta e \phi$, $\varepsilon(\xx)  =  \epsilon(\xx)/\epsilon\ind{w}$, $s(\xx)  =  -\sigma(\xx)/\sigma_0$. One can also introduce a rescaled polarizability defined as
\begin{equation}
\tilde \alpha = \frac{\beta }{(\beta q e_0 l\ind{GC})^2} \alpha.
\end{equation}
Usually instead of using the excess polarizability one can use the dielectric decrement $\tilde\beta$ in units of inverse Mole per liter \cite{Ben-Yaakov2011}, defined as $\alpha = \epsilon_0 \tilde\beta$. Typically the dielectric decrement for various salts is negative. 

The dimensionless polarizability represents an additional independent parameter of the theory. Finally we define the dimensionless fugacity as $ \tilde \lambda=2\pi\Xi l\ind{GC}^3\lambda$. 

We can estimate the numerical values for all these parameters and obtain typical values for monovalent counterions that are of the order: $l\ind{B}  \simeq  1\textrm{nm}$, $ \Xi  \simeq  1$, $\tilde\alpha  \simeq 10^{-2}$ and $\varepsilon\ind{ext}  \simeq 5\times 10^{-2}$.

We can now derive the grand canonical partition function in the form
\begin{equation}
\label{dl_pf}
\mZ=\int \exp\left(-\frac{S[\phi]}{\Xi}\right)[d\phi],
\end{equation}
where the field action can be obtained as
\begin{equation}
\label{dl_action}
S[\phi]= \frac{1}{8\pi}\int_E\varepsilon(\xx)(\nabla\phi(\xx))^2 d\xx - \frac{\lambda}{2\pi}\int_V \exp\left(-\frac{\alpha}{2}(\nabla\phi(\xx))^2+i\phi(\xx)\right) d\xx + \frac{i}{2\pi}\int_{\partial V}s(\xx)  \phi(\xx)d'\xx,
\end{equation}
Here, in order not to proliferate the notation we simply renamed all the "$\;\tilde {}\;$" quantities back to their un-"$\;\tilde {}\;$" symbols, because in what follows we will work only with the dimensionless action.
This expression is then the point of departure for the evaluation of the free energy and pressure of the system. One should note here that the partition function (\ref{dl_pf}) depends on two parameters: the coupling constant $\Xi$ as well as the dimensionless polarizability $\alpha$, \emph{i.e.} it is a two-parameter function.

\subsection{Density and electroneutrality}\label{}

In unscreened systems with long range Coulomb interactions the stability is insured only if the system as a whole is electroneutral. This is a particularity of long range interactions that becomes irrelevant for all finite range interaction potentials \cite{Kanduc2011}. Special care then needs to be taken in order to stipulate this stability, that is given as a 
condition on the one-particle ionic density. The latter is defined by the operator 
\begin{equation}
n(\xx)= \lambda \exp\left(-\frac{\alpha}{2}(\nabla\phi(\xx))^2+i\phi(\xx)\right)\mathbf{1}_V(\xx),
\end{equation}
where $\mathbf{1}_V(\xx)$ is the indicator function of the volume $V$ defined by $\int f(\xx)\mathbf{1}_V(\xx)d\xx=\int_V f(\xx)d\xx$. We will also use the indicator function of the surface $\partial V$, given by $\int f(\xx)\mathbf{1}_{\partial V}(\xx)d\xx=\int_{\partial V}f(\xx)d'\xx$.

While the true density is actually given by $\frac{n}{2\pi\Xi}$, the above expression is easier to use in the mean field approximation since it does not involve $\Xi$. 
We now impose average electroneutrality in the system by stipulating that 
\begin{equation}\label{electroneutrality}
\int \langle n(\xx)\rangle d\xx=\int_{\partial V}s(\xx)d'\xx.
\end{equation}
This zero-moment gauge condition insures that the system remains stable for any configuration of the charges. Electroneutrality needs to be formulated as an additional condition on the density function only for unscreened interactions, see \cite{Kanduc2011} for details.

\subsection{Grand potential, free energy, and pressure}\label{presdisc}

The grand canonical thermodynamic potential is defined by
\begin{equation}
J_{\lambda}=-\ln\mZ_{\lambda}.
\end{equation}
We write explicitly the dependence on $\lambda$ since it will feature prominently in our analysis. The fugacity is not a physical parameter, so the pressure should not depend on it. To solve this issue, we have to know how $J_\lambda$ depends on $\lambda$; by differentiating (\ref{dl_pf}), we get
\begin{equation}
\frac{dJ_{\lambda}}{d\lambda}=-\frac{N}{2\pi\Xi\lambda},
\end{equation}
where $N$ has no subscript $\lambda$ because it does not depend on it as a consequence of electroneutrality (\ref{electroneutrality}). Now, it is clear that the free energy defined by
\begin{equation}\label{def_free_en}
F_{\lambda}=J_{\lambda} + \frac{N}{2\pi\Xi}\ln\lambda
\end{equation}
does not depend on $\lambda$, \emph{i.e.} $F_{\lambda} = F$, while $\ln\lambda$ is the chemical potential. This means that the free energy can be safely used to compute the pressure:
\begin{equation}
P=-\frac{\partial F}{\partial L}.
\label{defpress1}
\end{equation}
Note that all the energies defined above are energies per unit area because of the transverse extensivity of our system. Furthermore the above pressure is in dimensionless units; the physical pressure is thus $p=P/(\beta l\ind{GC}^3)$.

\section{Weak coupling approximation}

Depending on the strength of the Coulomb coupling as parameterized by the coupling constant $\Xi$ the grand canonical partition function exhibits two well defined limiting laws \cite{Boroudjerdi2005,Kanduc2009}. For vanishing values of the coupling constant, $\Xi \rightarrow 0$,  the partition function can be well approximated by its saddle-point value and fluctuations around it. In fact the saddle-point is known to correspond exactly to the mean-field Poisson-Boltzmann expression while the Gaussian fluctuations around the mean-field correspond to the zero Matsubara frequency van der Waals or thermal Casimir interactions \cite{Podgornik1988}.

We will first derive the mean-field equations for our field-action, equivalent to those derived elsewhere \cite{Ben-Yaakov2011b,Frydel2011}, and then evaluate the Gaussian fluctuations around the mean-field and their dependence on the separation between the bounding surfaces.

\subsection{Mean-field}

We start with the general saddle-point equation satisfied at equilibrium
\begin{equation}\label{mf_eq}
0=\left\langle \frac{\delta S}{\delta\phi}\right\rangle=\frac{1}{2\pi}\left[-\nabla\cdot\left\langle\left[\frac{\varepsilon(\xx)}{2}+\alpha n(\xx)\right]\nabla\phi(\xx)\right\rangle-i\langle n(\xx)\rangle + i s(\xx)\mathbf{1}_{\partial V}(\xx)\right].
\end{equation}
The mean-field is more often written in terms of the (real) electrostatic potential proportional to $\psi=-i\phi$, with the corresponding field-action $\tilde S[\psi]$, than in terms of the fluctuating potential $\phi$. For this new variable,
the  mean-field configuration is evaluated from the saddle-point condition
\begin{equation}
\frac{\delta \tilde S[\psi_\textrm{MF}]}{\delta\psi(\xx)}=0.
\end{equation}
The grand canonical potential is then approximated by
\begin{equation}
J\simeq J_\textrm{MF}=\frac{\tilde S[\psi_\textrm{MF}]}{\Xi}.
\end{equation}
From the saddle-point equation the mean-field equation can be rewritten in its Poisson-Boltzmann form as \cite{Ben-Yaakov2011b,Frydel2011}
\begin{equation}\label{mf_bulk}
\nabla\cdot\left[\left(\frac{\varepsilon}{2} + \alpha n\right)\nabla\psi_\textrm{MF}\right]=- n + s \mathbf{1}_{\partial V},
\end{equation} 
where the density is given by
\begin{equation}\label{def_n}
n= \lambda \exp\left(\frac{\alpha}{2}(\nabla\psi_\textrm{MF})^2-\psi_\textrm{MF}\right)\mathbf{1}_V.
\end{equation}
In these two equations, it is clear that the fugacity can be absorbed into the electrostatic potential: this change will modify the grand potential but not the free energy. We can thus assume $\lambda=1$ for the mean-field as well as for fluctuations around it.

\subsection{Pressure in the plane-parallel geometry}

In 1D, which is also the case of two charged plane parallel surfaces since the mean potential depends only on the transverse coordinate $z$, the Poisson-Boltzmann equation has the form
\begin{equation}\label{mf_bulk1}
\left[\left(\frac{\varepsilon}{2} + \alpha n(z)\right)\psi_\textrm{MF}(z)'\right]'=-n(z) + s \mathbf{1}_{\partial V}(z),
\end{equation} 
with
\begin{equation}
n(z)= \lambda \exp\left(\frac{\alpha}{2}\psi_\textrm{MF}(z)'^2- \psi_\textrm{MF}(z)\right)\mathbf{1}_V(z).
\end{equation}
We used the notation $f'(z) = \frac{df}{dz}(z)$. In this case it can be shown that the pressure in the system is a constant given by the contact value theorem \cite{Ben-Yaakov2011}
\begin{equation}
P=\frac{1}{2\pi\Xi}\left[ n(z)-\left(\frac{\varepsilon}{4}+\alpha n(z)\right)\psi\ind{MF}'(z)^2 \right]=\text{const}.
\label{presform}
\end{equation}
It can be easily  checked that this quantity is actually equal to the pressure obtained equivalently by the standard thermodynamic definition $P=-\frac{1}{\Xi}\frac{\partial \tilde S[\psi\ind{MF}]}{\partial L}$. The above form for the interaction pressure contains an osmotic van't Hoff term, the first one in Eq. \ref{presform}, that contains the effects of the polarizability implicitly, \emph{i.e.} through the variation of the density profile on the polarizability, and a Maxwell stress term, the second one in Eq. \ref{presform}, that contains the polarizability effects explicitly.

\subsection{Second order fluctuations correction}

The grand potential can be computed to the next order by taking into account fluctuations around the mean-field solution. This is done by expanding $S$ around $\phi\ind{MF}=i\psi_\textrm{MF}$ to the second order, obtaining  
\begin{equation}
 S[\phi_\textrm{MF}+\theta] = \tilde S[\psi_\textrm{MF}] + \frac{1}{2}\int \frac{\delta^2 S}{\delta\phi(x)\delta\phi(y)}[\phi_\textrm{MF}] \theta(x)\theta(y)dx dy=\tilde S[\psi_\textrm{MF}]+S^{(2)}[\theta]. 
\end{equation}
In this case the grand potential is given by
\begin{equation}
J\simeq J_\textrm{MF}^{(1)}=\frac{\tilde S[\psi_\textrm{MF}]}{\Xi}-\ln\mZ^{(2)} = \frac{\tilde S[\psi_\textrm{MF}]}{\Xi}-\ln\left[\int\exp\left(-\frac{S^{(2)}[\theta]}{\Xi}\right)[d\theta]\right],
\end{equation}
where $\mZ^{(2)}$ is the contribution of the fluctuations to the partition function.
The effective action for the fluctuations $S^{(2)}[\theta]$ is straightforward to compute, yielding
\begin{eqnarray} 
S^{(2)}[\theta]&=&
\frac{1}{4\pi}\int\left[\left(\frac{\varepsilon}{2}+\alpha n\right)(\nabla \theta)^2+ n \alpha^2(\nabla\psi_\textrm{MF}\cdot\nabla \theta)^2 -  \left(\frac{1}{2}\nabla\cdot(\varepsilon\nabla\psi_\textrm{MF}) - s\mathbf{1}_{\partial V}\right)\theta^2\right],
\label{act_fluc}
\end{eqnarray}
where we have used the mean-field Eq. (\ref{mf_bulk}). 

Note that $\varepsilon\nabla\psi_\textrm{MF}$ is not continuous, and thus leads to a surface term.  As expected, this action does not depend on the fugacity but on the mean-field density, which is the only physically meaningful quantity. 

In our model we consider parallel plates of constant surface charge, the mean field problem is then one dimensional. We can therefore split the coordinates in a one dimensional coordinate $z$ perpendicular to the plates, and a two dimensional one parallel to the plates: $\xx=(z,\rr)$.

\subsection{Pressure}

Fourier transforming the fluctuations in the direction parallel to the plates we obtain
\begin{equation}
\theta(z,\rr)=\int \exp(i\kk\cdot\rr)\tilde\theta(z,\kk)\frac{d\kk}{(2\pi)^2},
\end{equation}
where $\tilde\theta(z,-\kk)=\tilde\theta(z,\kk)^*$ because $\theta$ is real. This decomposition furthermore allows us to write the fluctuations action (\ref{act_fluc}) as
\begin{equation}
S^{(2)}[\theta]=\int S_\kk^{(2)}\left[\tilde\theta(\cdot,\kk)\right]\frac{d\kk}{(2\pi)^2},
\end{equation}
where the one dimensional action is
\begin{eqnarray}
\label{act_fluc_k}	
S_\kk^{(2)}[\theta]&=&\frac{1}{4\pi}\int\left(\left[\frac{\varepsilon}{2}+\alpha n+ \alpha^2 n\psi_\textrm{MF}'^2\right]\theta'^2+\left[-\frac{1}{2}(\varepsilon \psi_\textrm{MF}')'+\left(\frac{\varepsilon}{2}+\alpha n\right) k^2\right]\theta^2\right)  \nonumber \\
&&\quad+\frac{1}{4\pi} \left[\theta(0)^2+\theta(L)^2\right] \nonumber\\
&=& S^{(2)}_{\kk,\textrm{b}}+S^{(2)}_{\kk,\textrm{s}}.
\end{eqnarray}
This action thus has a bulk part $S^{(2)}_{\kk,\textrm{b}}$ and a surface part $S^{(2)}_{\kk,\textrm{s}}$. The surface action actually contains another term due to the fact that $\varepsilon \psi_\textrm{MF}'$ is discontinuous across the bounding surfaces, so that finally
\begin{equation}
S^{(2)}_{\kk,\textrm{s}}[\theta]=\frac{C}{2}\left(\theta(0)^2+\theta(L)^2\right), \qquad {\rm where} \qquad C=\frac{1}{4\pi}\left[\varepsilon\psi\ind{MF}'\right]_{0^+}^{0^-}+\frac{1}{2\pi},
\end{equation}
with the notation $[f(x)]_{x_1}^{x_2}=f(x_2)-f(x_1)$. We also used the symmetry $z\leftrightarrow L-z$ of our system. The partition function for the fluctuations can be written as a product of path-integrals,
\begin{equation}\label{pi_prod}
\mZ^{(2)}=\prod_\kk \int \exp\left(- \frac{S^{(2)}_\kk[\theta]}{\Xi}	\right)[d\theta].
\end{equation}

These path-integrals are computed in appendix \ref{ap1}, leading to
\begin{equation}
\mZ^{(2)}_\kk=\exp \left(\frac{kL}{2}\right)\sqrt{\frac{2\pi b^\kk(0,L)}{\left[a\ind{f}^\kk(0,L)+\frac{ C+\deltext k/4\pi}{\Xi}\right]^2-b^\kk(0,L)^2}},
\end{equation}
where the functions $b^\kk$ and $a\ind{f}^\kk$ are defined in the appendix. 

The total free energy of the mean field configuration and fluctuations around it is then obtained as
\begin{equation}
F\ind{MF}^{(1)}  =  \frac{\tilde S[\psi\ind{MF}]}{\Xi}-\frac{1}{2\pi\beta}\int_0^\infty \ln\left(\mZ^{(2)}_k\right)k\,dk.
\label{pressure1}
\end{equation}
We note here that the structure of the free energy $F\ind{MF}^{(1)} $ does not look like a mean-field term independent of  the counterion polarizability plus a zero frequency van der Waals term that stems from the polarizability of the counterions. Though this kind of decomposition is sometimes assumed  in the literature \cite{Ninham1997,Edwards2004}, it clearly does not correspond to the weak-coupling approximation. 

\begin{figure*}[t!]
\begin{center}
	\begin{minipage}[b]{0.485\textwidth}
	\begin{center}
		\includegraphics[width=\textwidth]{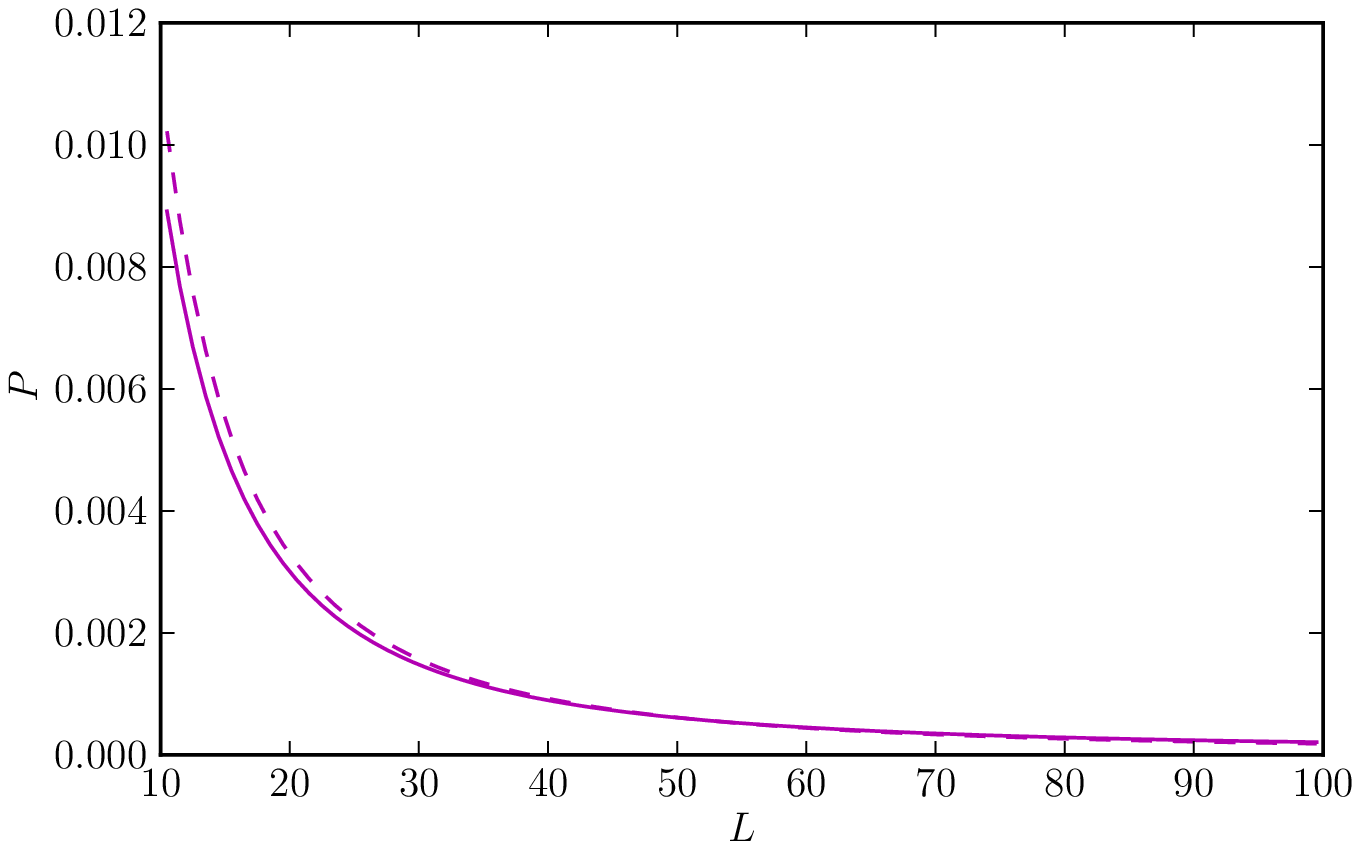}
	\end{center}\end{minipage} \hskip0.25cm
	\begin{minipage}[b]{0.485\textwidth}\begin{center}
		\includegraphics[width=\textwidth]{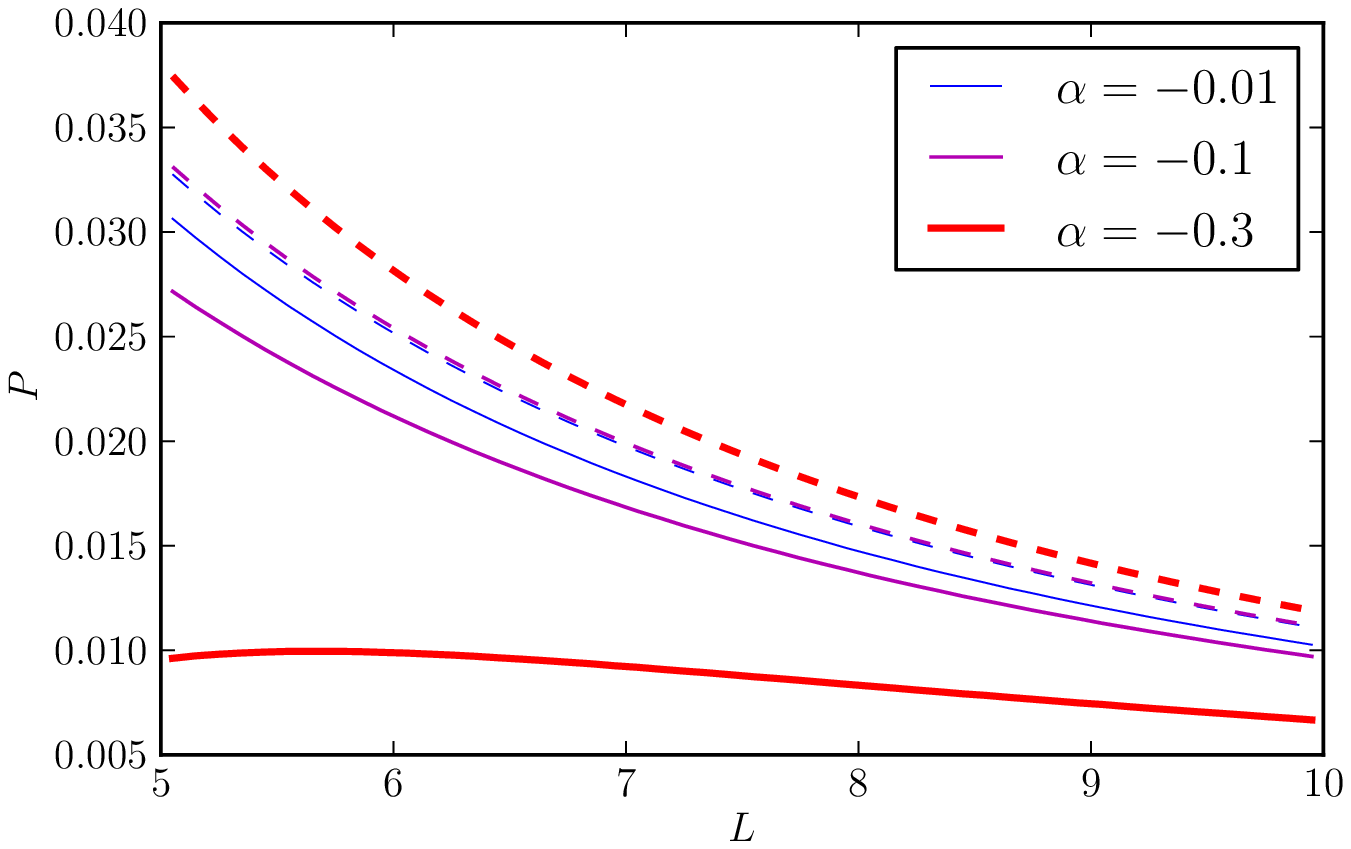}
	\end{center}\end{minipage} \hskip0.25cm	
\caption{Pressure $P^{(1)}(L)$ from Eq. \ref{defpress2} as a function of the plate separation $L$ for $\varepsilon\ind{ext}=0.05$, $\Xi=1$, and $R=1$. The dashed lines are the mean field result, the solid lines include the fluctuations.
\emph{Left}: for $\alpha=-0.1$ and large plate separation,  the difference is barely distinguishable.
\emph{Right}: for small plate separation $L$ and various values of the polarizability $\alpha$. The effect of fluctuations can  be quite important for large counterion polarizabilities.}
\label{f_L_P}
\end{center}\end{figure*}

We can see numerically that the  integral over the transverse Fourier modes in Eq. (\ref{pressure1})  diverges; this comes from our model of point-like dipoles. Taking into account the size $R$ of the polarizable ions (more precisely, $R$ is the charge of their hydration shell), the integral is regularized by the dimensionless cut-off
\begin{equation}
k\ind{max}=\frac{\pi l\ind{GC}}{R}.
\end{equation}
Physically the cut-off arises because electric fields which fluctuate on length scales shorter than the 
polarizable ion cannot polarize it. 
The interaction pressure on this level of approximation is then obtained by taking into account Eq. \ref{defpress1}, leading to
\begin{equation}
P^{(1)}=-\frac{\partial F^{(1)}\ind{MF}}{\partial L}.
\label{defpress2}
\end{equation}
The results for the fluctuations-corrected interaction pressure from Eq. \ref{defpress2} on the weak coupling approximation level are shown on Fig. \ref{f_L_P} for $\Xi=1$, $\varepsilon\ind{ext}=0.05$ and $R=1$, for various values of the counterion polarizability $\alpha$. The fluctuations correction in $P^{(1)}$ is quite small compared to the mean-field value, but can become substantial as the polarizability $\alpha$ decreases, \emph{i.e.} becomes more negative. This correction reduces the interaction pressure between the surfaces. This indicates that ions with nominally equal charge (of equal valency) but differing in the polarizability will mediate markedly different interactions when confined between charged dielectric interfaces even at the weak-coupling level.  

\subsection{Density}\label{}

We now consider  the ion density by taking into account the mean field solution as well as the fluctuations around the mean field. From (\ref{def_n}), the ion density is
\begin{eqnarray}
\rho_1(\xx) & = & \left\langle \exp\left(-\frac{\alpha}{2}(\nabla\phi(\xx))^2+i\phi(\xx)\right)\right\rangle_1 \\
& = & (2\pi\alpha)^{-3/2}\int \exp\left(-\frac{p^2}{2\alpha}\right)\left\langle \exp\left(i\pp\cdot\nabla\phi(\xx)+i\phi(\xx)\right)\right\rangle_1 d\pp,
\end{eqnarray}
where we used a Hubbard-Stratonovitch transformation to obtain the last expression. In this way we have only terms linear in $\phi$ in the exponential. The subscript 1 denotes that we take into account the first order of the fluctuations. We notice that the mean-field equation for electroneutrality should also hold on average at equilibrium, so that $\rho_1$ will satisfy electroneutrality. As a consequence, we only need $\rho_1$ up to a multiplicative constant, and this constant will be set by electroneutrality.

The interpretation of the above formula is that the local ion density is the average over a fluctuating dipolar moment vector of a Coulomb fluid characterized by ions with a charge and a dipolar moment. We then decompose $\phi$ into a mean-field term plus Gaussian fluctuations
\begin{equation}
\phi=i\psi\ind{MF}+\theta,
\end{equation}
obtaining
\begin{equation}\label{rho1_p}
\rho_1(\xx)=(2\pi\alpha)^{-3/2}\int \exp\left(-\frac{p^2}{2\alpha}-\pp\cdot\nabla\psi\ind{MF}(\xx)-\psi\ind{MF}(\xx)\right)\left\langle \exp\left(i\pp\cdot\nabla\theta(\xx)+i\theta(\xx)\right)\right\rangle_1 d\pp.
\end{equation}
The average is now easy to compute,
\begin{equation}
\left\langle \exp\left(i\pp\cdot\nabla\theta(\xx)+i\theta(\xx)\right)\right\rangle_1=\exp\left(-\frac{1}{2}\left\langle(\pp\cdot\nabla\theta(\xx)+\theta(\xx))^2\right\rangle_1\right),
\end{equation}
and then, introducing the correlator of the fluctuations
\begin{equation}
G(\xx,\xx')=\langle\theta(\xx)\theta(\xx')\rangle_1, 
\end{equation}
we can write it as
\begin{equation}
\left\langle \exp\left(i\pp\cdot\nabla\theta(\xx)+i\theta(\xx)\right)\right\rangle_1=\exp\left(-\frac{1}{2}\pp^T\nabla\nabla'^TG(\xx,\xx)\pp-\frac{1}{2}G(\xx,\xx)-\frac{1}{2}\pp\cdot\bar\nabla G(\xx,\xx)\right).
\end{equation}
We used the notation $\nabla$ for the gradient with respect to the first variable of $G(\xx,\xx')$, $\nabla'$ for the second variable, and $\bar\nabla$ for the sum of the two gradients. We can now insert this expression into (\ref{rho1_p}), and remain with a Gaussian integral
\begin{equation}
\rho_1(\xx)=(2\pi\alpha)^{-3/2}\int \exp\left(-\frac{1}{2}\pp^T\aal_{1+}^{-1}(\xx)\pp-\pp\cdot\nabla\psi_1(\xx)-\psi_1(\xx)\right)d\pp,
\end{equation}
where we introduced a renormalized polarizability (which is now a position-dependent matrix) and a renormalized field
\begin{eqnarray}
\aal_{1+}^{-1}(\xx) & = & \alpha^{-1}+\nabla\nabla'^T G(\xx,\xx),\\
\psi_1(\xx) & = & \psi\ind{MF}(\xx) + \frac{1}{2}G(\xx,\xx).
\end{eqnarray}
Performing the integral gives
\begin{equation}\label{rho1+}
\rho_{1+}(\xx) = \sqrt{\det\left(\frac{\aal_{1+}(\xx)}{\alpha}\right)}\exp\left(\frac{1}{2}[\nabla\psi_1(\xx)]^T\aal_{1+}(\xx)\nabla\psi_1(\xx)-\psi_1(\xx)\right).
\end{equation}
The index "$+$" means that our computation works only for $\alpha>0$. In the more common case where $\alpha<0$, the computation is the same up to some factors of $i$, and we get
\begin{equation}
\aal_{1-}^{-1}(\xx) = |\alpha|^{-1}-\nabla\nabla'^T G(\xx,\xx),
\end{equation}
and 
\begin{equation}
\rho_{1-}(\xx)= \sqrt{\det\left(\frac{\aal_{1-}(\xx)}{\alpha}\right)}\exp\left(-\frac{1}{2}[\nabla\psi_1(\xx)]^T\aal_{1-}(\xx)\nabla\psi_1(\xx)-\psi_1(\xx)\right).
\end{equation}

Now we have to compute $G(\xx,\xx)$ and $\nabla\nabla'^T G(\xx,\xx)$ at each point. To do this, we will use the same technique  we used to compute the pressure: we Fourier transform the fluctuations in the direction parallel to the plates and use the Pauli-van Vleck formula.

Since the fluctuations action (\ref{act_fluc}) is a sum over different transversal modes, two modes with different wave vectors are uncorrelated and we can write the correlator as an integral over the modes,
\begin{equation}
 G(\xx,\xx') = \int \exp(i\kk\cdot[\rr-\rr'])G_\kk(z,z')\frac{d\kk}{(2\pi)^{d-1}},
\end{equation}
where $G_\kk(z,z')$ is the one dimensional correlator for the action in Eq.  (\ref{act_fluc_k}). More precisely, we need $G_\kk(z,z)$ as well as $\partial\partial'G_\kk(z,z)$. These functions are computed in appendix \ref{ap2}. Then we can write
\begin{equation}
G(\xx,\xx)=\int G_\kk(z,z)\frac{d\kk}{(2\pi)^{2}}=\frac{1}{2\pi}\int_0^\infty G_k(z,z)k\,dk,
\end{equation}
and the matrix
\begin{equation}
\nabla\nabla'^TG(\xx,\xx)=\int \begin{pmatrix}\partial\partial' & -i\kk^T\partial \\ i\kk\partial' & \kk\kk^T\end{pmatrix}G_\kk(z,z)\frac{d\kk}{(2\pi)^2}=\frac{1}{(2\pi)}\int_0^\infty \begin{pmatrix}\partial\partial' & 0 \\ 0 & \frac{k^2}{2}\mathbf{1}_2\end{pmatrix}G_k(z,z) k\,dk,
\end{equation}
where $\mathbf{1}_2$ is the two dimensional identity matrix.

In conclusion, we have the algorithm to compute the pressure and the density: for each mode, we integrate the Pauli-van Vleck formula and then compute its contributions to $G(\xx,\xx)$ and $\nabla\nabla'^TG(\xx,\xx)$ and add them to the contributions of the previous modes. Finally, we compute the new density at each point and renormalise it using electroneutrality.

\begin{figure*}[t!]
\begin{center} 
	\begin{minipage}[b]{0.485\textwidth}
	\begin{center}
		\includegraphics[width=\textwidth]{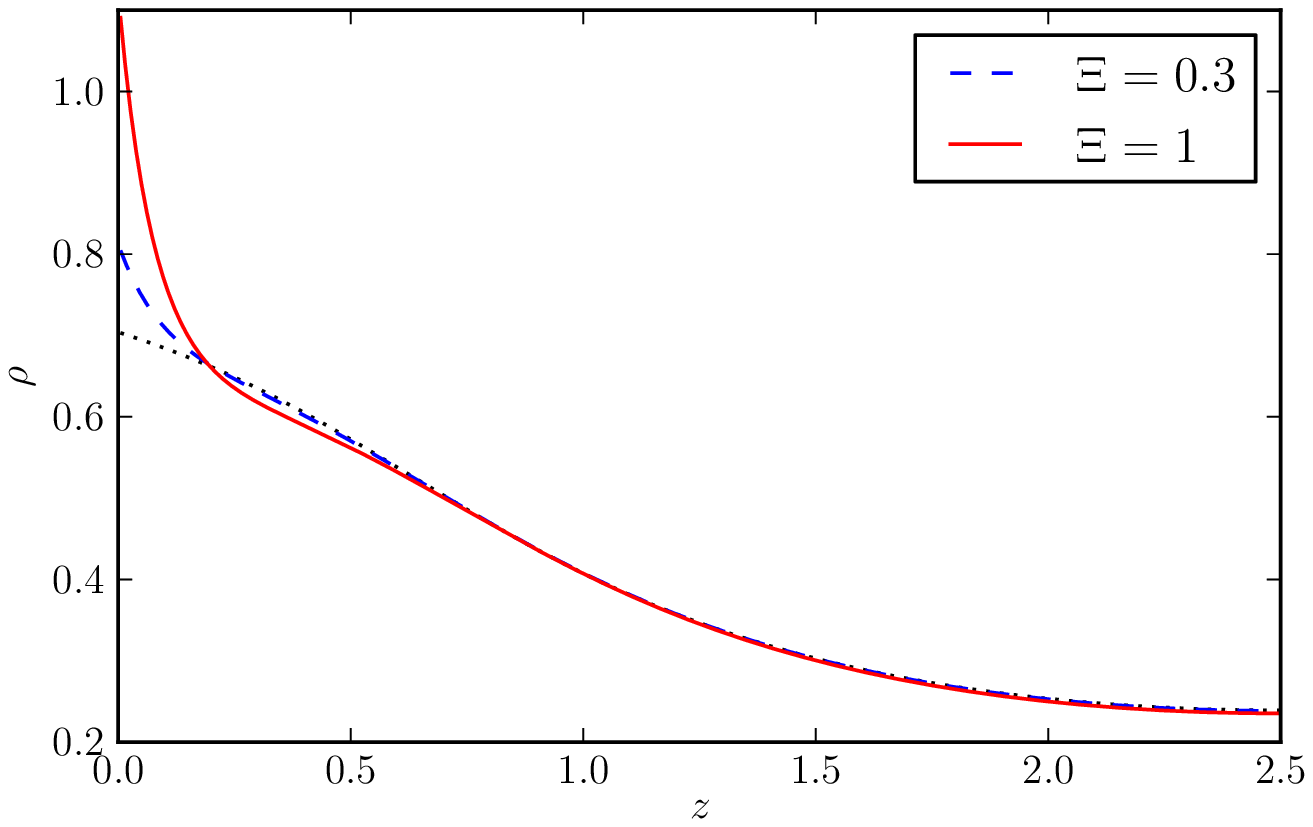}
	\end{center}\end{minipage} \hskip0.25cm
	\begin{minipage}[b]{0.485\textwidth}\begin{center}
		\includegraphics[width=\textwidth]{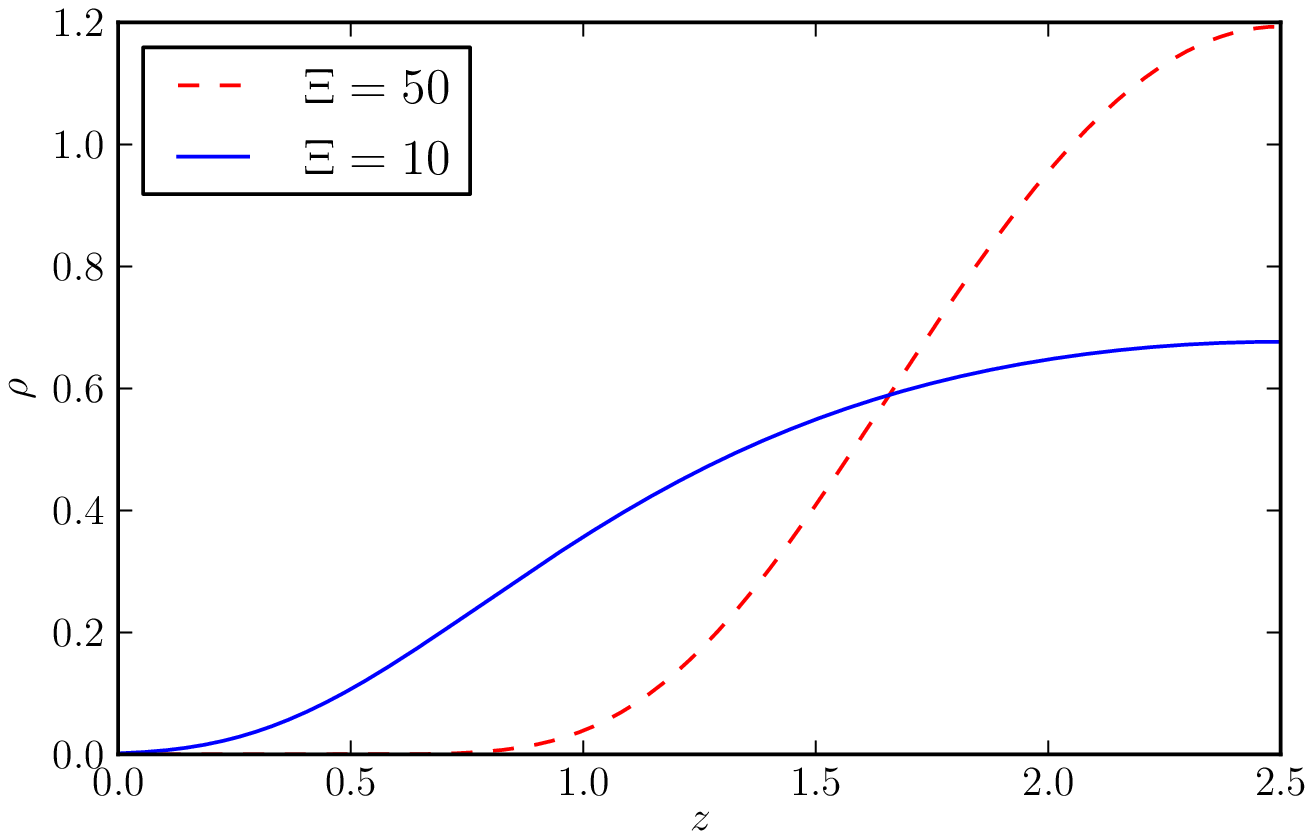}
	\end{center}\end{minipage} \hskip0.25cm	
\caption{Counterions density profile within the slab - dependence on the coupling constant $\Xi$. \emph{Left}: Weak coupling density close to the left electrode taking into account the fluctuations around the mean field as a function of the position within the slab ($z\in\left[0,\frac{L}{2}\right]$), for $R=0.3$, $\alpha=-0.3$, $\varepsilon\ind{ext}=0.05$ and $\Xi=0.3$ (dashed line) or $\Xi=1$ (solid line). The mean-field density itself is presented by the dotted line. \emph{Right}: Strong coupling density as a function of the position for $R=2$, $\alpha=-0.01$, $\varepsilon\ind{ext}=0.05$ and $\Xi=10$ (solid line) or $\Xi=50$ (dashed line). 
}
\label{f_z_rho}
\end{center}\end{figure*}

\begin{figure*}[t!]
\begin{center} 
	\begin{minipage}[b]{0.485\textwidth}
	\begin{center}
		\includegraphics[width=\textwidth]{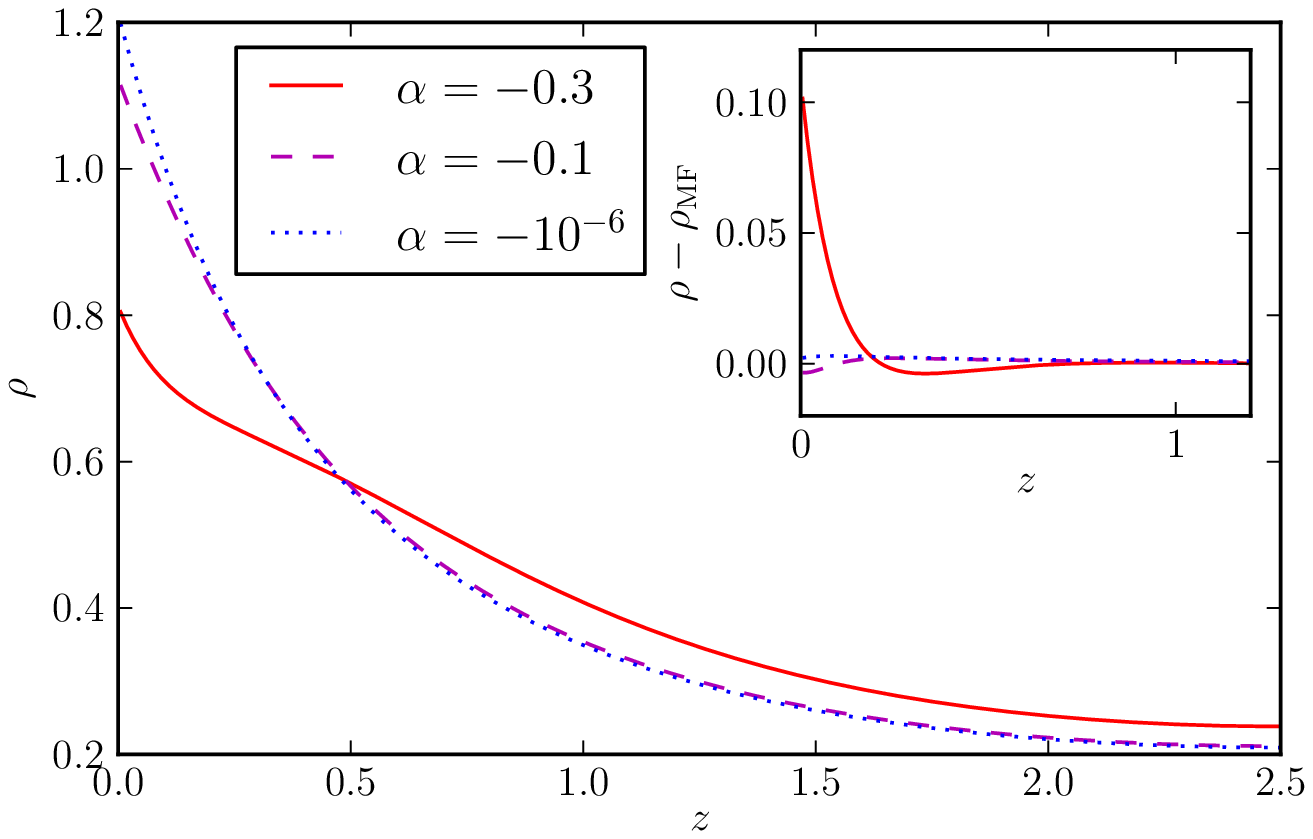}
	\end{center}\end{minipage} \hskip0.25cm
	\begin{minipage}[b]{0.485\textwidth}\begin{center}
		\includegraphics[width=\textwidth]{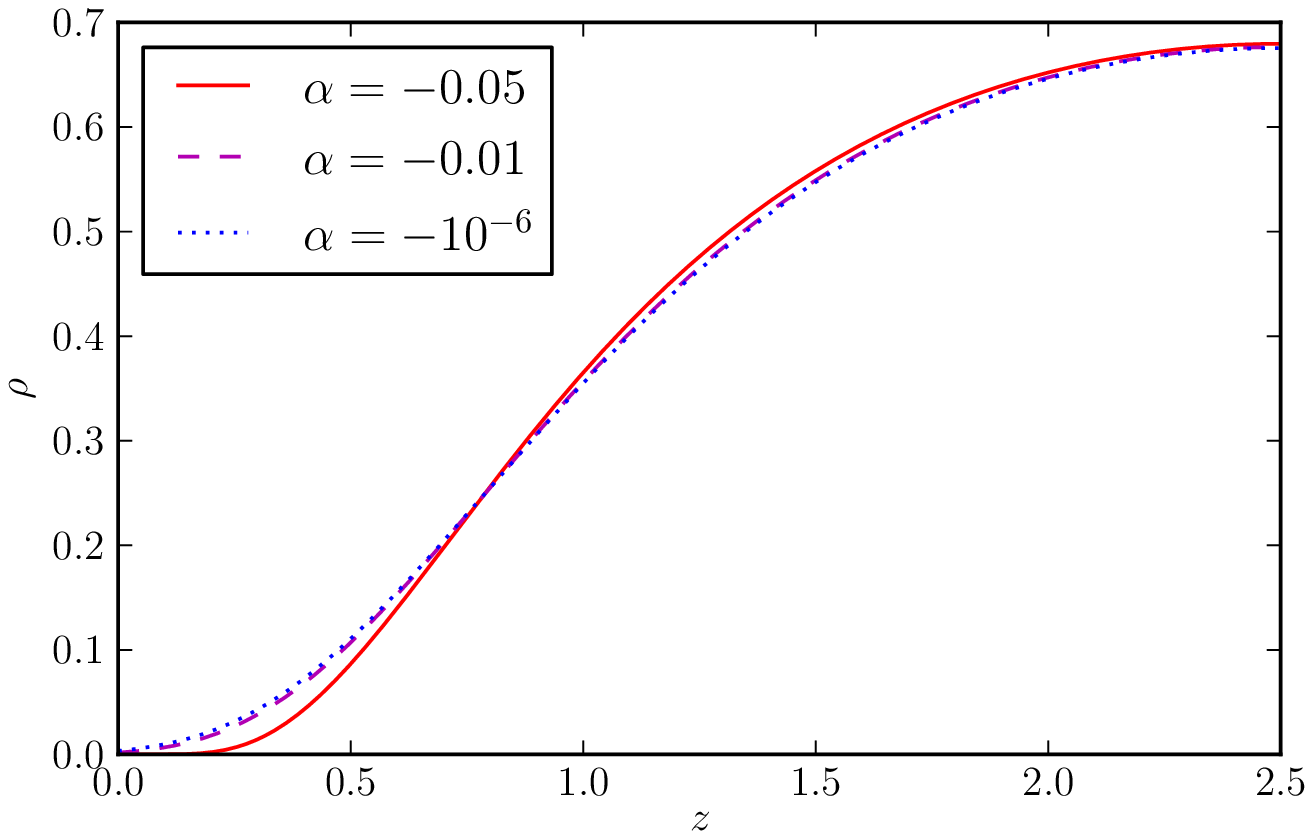}
	\end{center}\end{minipage} \hskip0.25cm	
\caption{Counterions density profile within the slab -- dependence on the polarizability $\alpha$. \emph{Left}: Weak coupling density close to the left electrode taking into account the fluctuations around the mean field as a function of the position within the slab ($z\in\left[0,\frac{L}{2}\right]$), for $R=0.3$, $\varepsilon\ind{ext}=0.05$, $\Xi=0.3$ and $\alpha=-0.3$ (solid line), $\alpha=-0.1$ (dashed line) or $\alpha=-10^{-6}$ (dotted line). \emph{Inset}: Deviation from the mean-field density. \emph{Right}: Strong coupling density as a function of the position within the dielectric slab for $R=1$, $\varepsilon\ind{ext}=0.05$, $\Xi=10$ and $\alpha=-0.05$ (solid line), $\alpha=-0.01$ (dashed line) or $\alpha=-10^{-6}$ (dotted line).
}
\label{f_z_rho_alpha}
\end{center}\end{figure*}

The mean field and first order densities can be compared on Fig. \ref{f_z_rho} (left). First of all, we observe that the effect of the fluctuations is small and depends on the values of the parameters. For higher $\Xi$, e.g. $\Xi=1$ on the figure, the ions get preferentially included in the region close to the dielectric boundaries.  This is not a mean-field effect since the mean-field density does not depend on the coupling parameter.
The $\alpha$ dependence of the counterion density in the slab is shown on Fig. \ref{f_z_rho_alpha} (left). The mean field density depends strongly on $\alpha$ \cite{Ben-Yaakov2011,Kanduc2009}, and this dependence remains after one adds the fluctuation contribution. The inset shows that the deviation from the mean-field density increases with $\alpha$: the effect of the fluctuations is enhanced by the polarizability.

As a conclusion, the counterions are attracted by the boundaries, and most of this effect has a mean-field nature. The polarizability tends to lower this attraction at the mean-field level, but to increase it at the fluctuations level.

\section{Strong coupling}

The strong coupling approximation \cite{Boroudjerdi2005,Kanduc2009} is obtained in the limit of asymptotically large coupling parameter, $\Xi \rightarrow \infty$. In this limit it turns out that the statistical mechanical description of the system is equivalent to a properly normalized one-body description. This means that we can treat the system as composed of bounding surfaces and a single polarizable charge between them. 
We will first derive the strong coupling form for the partition function, equivalent to the first order virial expansion, and then evaluate the density profile and the interaction pressure.

\subsection{Formulation}

The strong coupling limit formally corresponds to the $\lambda \ll 1$ limit. To the lowest non-trivial order in $\lambda$, the partition function is given by
\begin{equation}
\mZ \simeq Z_0+\frac{\lambda}{2\pi\Xi} Z_1=Z_0\left(1+\frac{\lambda}{2\pi\Xi} U\right).
\end{equation}
We are thus interested in the evaluation of 
\begin{equation}
Z_0 = \int[d\phi] \exp\left(-\frac{1}{2\pi\Xi}\left[\frac{1}{4}\int\varepsilon(\xx)(\nabla\phi(\xx))^2 d\xx + i\int_{\partial V} s(\xx)\phi(\xx)d'\xx\right]\right).
\end{equation}
and $Z_1=\int z_1(\xx_0)d\xx_0$, with

\begin{eqnarray}
z_1(\xx_0) & = & \int[d\phi] \exp\left(-\frac{1}{2\pi\Xi}\left[\frac{1}{4}\int\varepsilon(\xx)(\nabla\phi(\xx))^2 d\xx+i\int_{\partial V}s(\xx)\phi(\xx)d'\xx\right]\right. \label{SC_z1} \\
&&\phantom{\int[d\phi] \exp()}
\left.-\left[\frac{\alpha}{2}(\nabla\phi(\xx_0))^2-i\phi(\xx_0)\right]\right). \nonumber
\end{eqnarray}
The quantity $\lambda z_1(\xx_0)/\mZ\simeq \lambda z_1(\xx_0)/Z_0$ is the ionic density at $\xx_0$. The total number of ions thus follows by stipulating that $\lambda U=N$, so that $\lambda$ can be tuned to satisfy electroneutrality.

As in the mean-field approximation, we will be interested in the density and the pressure. For the density, we will be specifically interested in the $\xx_0$ dependent part of $z_1(\xx_0)$, whereas for the pressure we need the $L$ dependent part of $Z_0$ and $Z_1$. In this sense the density is easier to compute, so that we address this question first.

\subsection{Density}\label{}

We introduce an auxiliary vector $\pp$ together with a Hubbard-Stratonovich decomposition and perform the integration over $\phi$ to write down Eq. \ref{SC_z1}\ as 
\begin{eqnarray}
z_1(\xx_0) & = & (2\pi\alpha)^{-3/2}\det\left(-\frac{\varepsilon\nabla^2}{4\pi\Xi}\right)^{-1/2} \nonumber \\
 & & \times \int d\pp \exp\left(-\frac{\pp^2 }{2\alpha}-\frac{1}{2}\left\langle\left( -\frac{1}{2\pi\Xi}\int s(\xx)\phi(\xx)d'\xx+\pp\cdot\nabla\phi(\xx_0)+\phi(\xx_0)\right)^2\right\rangle_{0}\right),
\end{eqnarray}
where $\langle \dots\rangle_0$ denotes the Gaussian average over $\phi$ 
with the action $S_0[\phi]=\frac{1}{8\pi}\int\varepsilon(\xx)(\nabla\phi(\xx))^2 d\xx$.
In this way the average can be written in terms of the correlator, $G(\xx,\xx')=\langle\phi(\xx)\phi(\xx')\rangle_0$, as
\begin{equation}
\left\langle\left(-\frac{1}{2\pi\Xi}\int s(\xx)\phi(\xx)d'\xx+\pp\cdot\nabla\phi(\xx_0)+\phi(\xx_0)\right)^2\right\rangle_0=\pp^T\AA(\xx_0)\pp+2\pp\cdot\BB(\xx_0)+C(\xx_0),
\end{equation}
where
\begin{eqnarray}
\AA(\xx_0) & = & \nabla\nabla'^TG(\xx_0,\xx_0),\label{def_A}\\
\BB(\xx_0) & = & \nabla \left(G(\xx_0,\xx_0) - \frac{1}{2\pi\Xi}\int s(\xx)G(\xx_0,\xx)d'\xx\right), \\
C(\xx_0) & = & G(\xx_0,\xx_0) - \frac{1}{\pi\Xi}\int s(\xx)G(\xx_0,\xx)d'\xx+\frac{1}{(2\pi\Xi)^2}\int s(\xx)s(\xx')G(\xx,\xx')d'\xx d'\xx' \nonumber\\
&=& C'(\xx_0)+\frac{1}{(2\pi\Xi)^2}\int s(\xx)s(\xx')G(\xx,\xx')d'\xx d'\xx' ,
\end{eqnarray}
and $\nabla$ and $\nabla'$ denote respectively the gradient with respect to the first and second variable. We can now perform the explicit integration over $\pp$, obtaining
\begin{equation}\label{z1}
z_1(\xx_0) = Z_0 \det\left(1+\alpha\AA(\xx_0)\right)^{-1/2} \times \exp\left(\frac{1}{2}\BB(\xx_0)^T\left(\frac{1}{\alpha}+\AA(\xx_0)\right)^{-1}\BB(\xx_0)-\frac{C'(\xx_0)}{2}\right).
\end{equation}
where 
\begin{equation}\label{sc_z0}
Z_0 = \det\left(-\frac{\varepsilon\nabla^2}{4\pi\Xi}\right)^{-1/2}\times\exp\left(-\frac{1}{2(2\pi\Xi)^2}\int s(\xx)s(\xx')G(\xx,\xx')d'\xx d'\xx'\right).
\end{equation}

As we noted in the weak-coupling treatment, the Hubbard-Stratonovitch transform depends on the sign of $\alpha$. However, it is easy to see here that the final expressions (\ref{def_A}-\ref{z1}) remain the same if $\alpha$ is negative.

We should mention that a problem arises in Eq. (\ref{z1}) if an eigenvalue of $1+\alpha\AA(\xx_0)$ is negative. This happens notably if $\alpha$ is too negative, leading to a negative effective permittivity of the hydration shell of the ion, and thus to an instability for the field. The problematic value of $\alpha$ therefore strongly depends on the radius of the hydration shell.

\begin{figure*}[t!]
\begin{center}
	\begin{minipage}[b]{0.485\textwidth}
	\begin{center}
		\includegraphics[width=\textwidth]{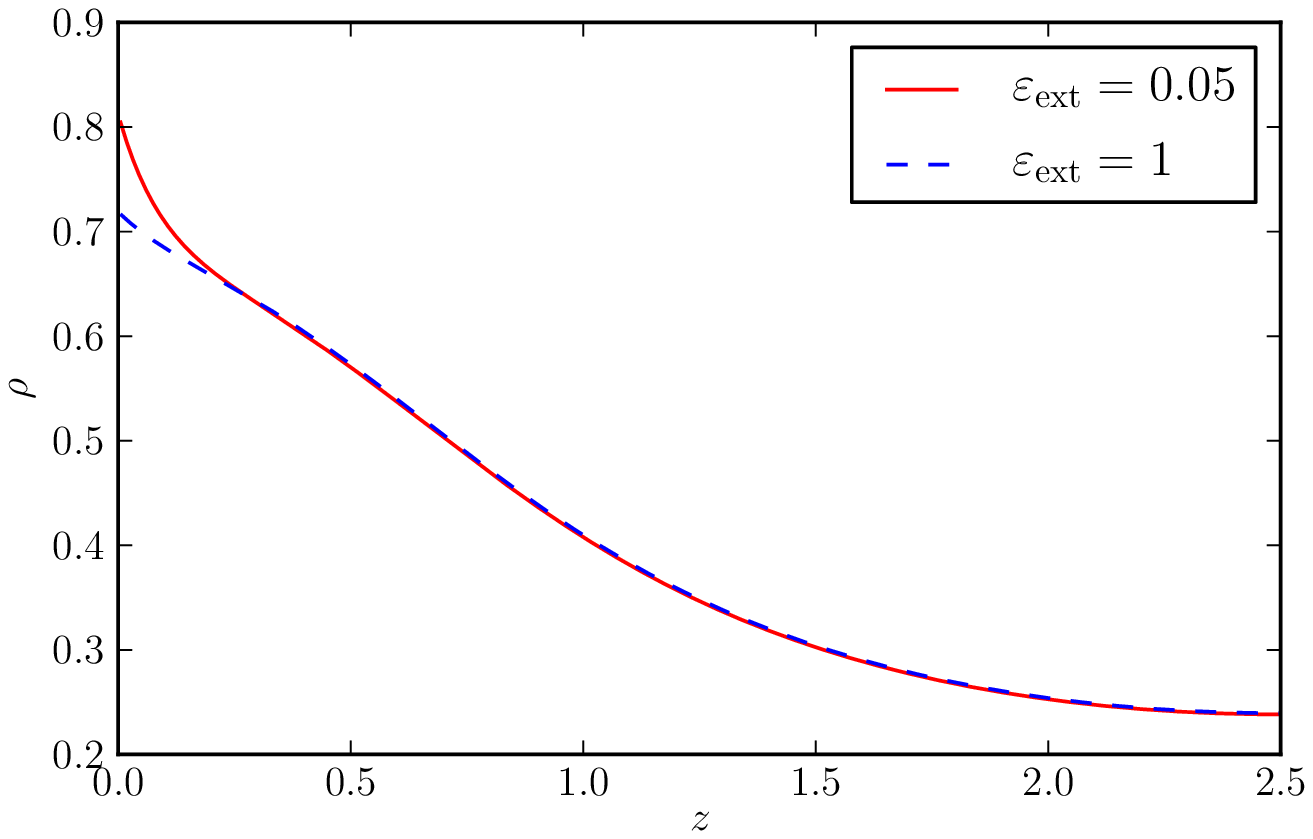}
	\end{center}\end{minipage} \hskip0.25cm
	\begin{minipage}[b]{0.485\textwidth}\begin{center}
		\includegraphics[width=\textwidth]{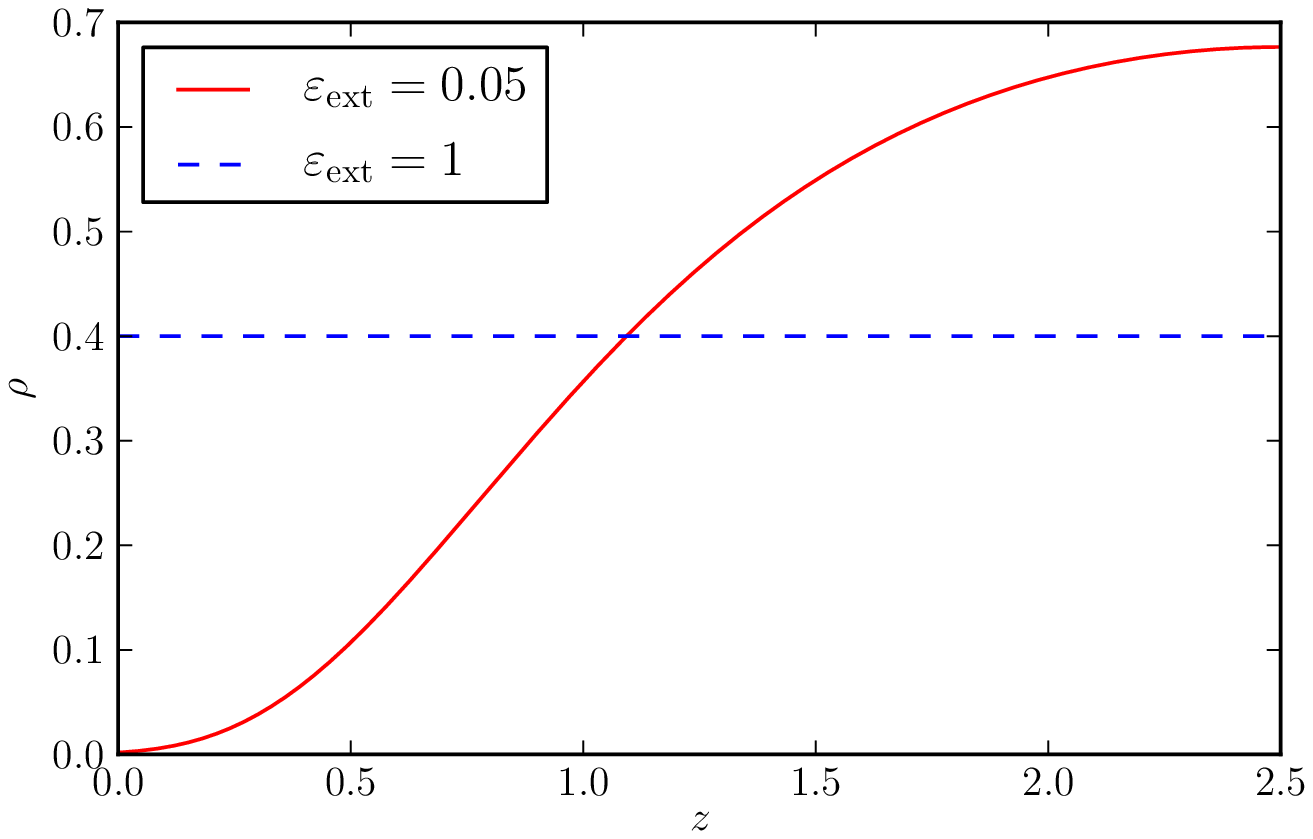}
	\end{center}\end{minipage} \hskip0.25cm	
\caption{Effect of the outer permittivity on the counterion density distribution. 
\emph{Left}: Weak coupling density with $\alpha=-0.3$, $R=0.3$, $\Xi=0.3$ and $\varepsilon\ind{ext}=0.05$ (solid line) or $\varepsilon\ind{ext}=1$ (dashed line).
\emph{Right}: Strong coupling density with $\alpha=-0.01$, $R=2$, $\Xi=10$ and $\varepsilon\ind{ext}=0.05$ (solid line) or $\varepsilon\ind{ext}=1$ (dashed line).
}
\label{f_z_rho_eps}
\end{center}\end{figure*}

In order to be more explicit, we need the expression for the correlator. Again, we Fourier transform the field in the direction parallel to the plates:
\begin{equation}
\phi(z,\rr)=\int \exp(i\kk\cdot\rr)\tilde\phi(z,\kk)\frac{d\kk}{(2\pi)^{d-1}},
\end{equation}
and the correlator for the $\kk$ mode is relatively easy to determine and is given in \cite{Kanduc2007}. To make the symmetry $z\rightarrow L-z$ more explicit, we switch to coordinates where the plates are located at $-L/2$ and $L/2$; in this case the correlator is given by
\begin{equation}\label{correl_k}
G_\kk(z,z')=4\pi\Xi\left[\frac{\exp(-k|z-z'|)}{2k}+ \frac{\cosh(k(z+z'))+\Delta \exp(-kL)\cosh(k(z-z'))}{\Delta^{-1}\exp(kL)-\Delta \exp(-kL)}\right],
\end{equation}
where
\begin{equation}
\Delta=\frac{1-\varepsilon\ind{ext}}{1+\varepsilon\ind{ext}}.
\end{equation}
We will write $\AA(\xx_0)$, $\BB(\xx_0)$ and $C'(\xx_0)$ using this expression for the correlator. Divergences may appear, but for the density itself we can drop (almost) all the $\xx_0$-independent terms. In fact we find
\begin{equation}\label{AA}
A(\xx_0) = \frac{1}{2\pi}\int_0^{k\ind{max}}\begin{pmatrix}\partial\partial' &0 \\ 0 & \frac{k^2}{2}\mathbf{1}_2 \end{pmatrix} G_k(z_0,z_0) k\, dk,
\end{equation}
where we need a cut-off as in the weak-coupling limit, and
\begin{equation}
\partial\partial'G_k(z_0,z_0) = 4\Xi\left[q\ind{max}(k)-k\arctan\left(\frac{q\ind{max}(k)}{k}\right)\right]+4\pi\Xi k^2\frac{\cosh(2kz_0)-\Delta \exp(-kL)}{\Delta^{-1}\exp(kL)-\Delta \exp(-kL)},
\end{equation}
where $q\ind{max}(k)$ is defined by (\ref{defqmax}), $q\ind{max}(k)^2+k^2=k\ind{max}^2$. Then, for $\BB(\xx_0)$, we will keep
\begin{equation}\label{BB}
\BB(\xx_0) = 2\Xi\begin{pmatrix}
1\\ \mathbf{0} \end{pmatrix} \int_0^{k\ind{max}}\frac{k^2\sinh(2kz_0)}{\Delta^{-1}\exp(kL)-\Delta \exp(-kL)}dk.
\end{equation}
and we can drop the second term in $C'(\xx_0)$,
\begin{equation}
C'(\xx_0) = 2\Xi\int_0^{k\ind{max}}\frac{\cosh(2kz_0)}{\Delta^{-1}\exp(kL)-\Delta \exp(-kL)}k\, dk.
\end{equation}
The result is shown in Fig. \ref{f_z_rho} (right),  where we used electroneutrality to normalize the strong coupling result, defined as we have seen up to a constant. 
We see that the counterions are completely excluded from the region close to the interfaces and pushed towards the middle of the dielectric slab. This effect increases with the coupling parameter $\Xi$. It appears on Fig. \ref{f_z_rho_alpha}\ that, in opposition to the weak coupling limit, the dependence on the polarizability is weak. Fig. \ref{f_z_rho_eps}\ shows that the strong coupling density is ruled by the images: without them, the density would be constant within the slab \cite{Boroudjerdi2005,Jho2008,Kanduc2009,Naji2005}.

As a conclusion, the polarizability has a small effect at the strong coupling level.

\subsection{Pressure}\label{}

Using its definition (\ref{def_free_en}) and the condition $\lambda U=N$, we can write the $L$ dependent part of the free energy, up to the first order in $\lambda$, as

\begin{equation}
F=J+\frac{N}{2\pi\Xi}\ln \lambda = -\ln\mZ-\frac{N}{2\pi\Xi}\ln U \simeq -\ln Z_0-\frac{N}{2\pi\Xi}\ln U.
\label{fren}
\end{equation}

Of course we need to take into account the zero-moment gauge condition (electroneutrality, Eq.\ref{electroneutrality}) when evaluating the above expression, which to the lowest order eliminates the $\Xi$ dependence. 

Let us first compute the $L$ dependent part of $J_0=-\ln Z_0$, using (\ref{sc_z0}) we get 

\begin{equation}
J_0=\frac{1}{2}\ln \left[\det\left(-\frac{\varepsilon\nabla^2}{4\pi\Xi}\right)\right]+\frac{1}{2(2\pi\Xi)^2}\int s(\xx)s(\xx')G(\xx,\xx')d'\xx d'\xx'.
\end{equation}
The first term is the thermal Casimir fluctuation free energy, and the second the electrostatic interaction between the plates. Using the decomposition of the correlator in orthogonal modes, we can write
\begin{equation}
\int s(\xx)s(\xx')G(\xx,\xx')d'\xx d'\xx'=2\left(G_0(0,0)+G_0(0,L)\right),
\end{equation}
where we have taken $\int d\rr=1$ to have the energy per unit area. Since the orthogonal mode $k=0$ is ill-defined in (\ref{correl_k}), we can derive with respect to $L$ before taking the limit $k\rightarrow 0$. We get
$\frac{d}{dL}G_k(0,0)  \underset{k\rightarrow 0}{\rightarrow}  0$ and $
\frac{d}{dL}G_k(0,L)  \underset{k\rightarrow 0}{\rightarrow}  -2\pi\Xi$, so that we can replace $G_0(0,L)=-2\pi\Xi L$. We thus get the final expression for the grand potential, \begin{equation}
J_0 =\frac{1}{4\pi}\int \ln\left(1-\Delta^2\exp(-2kL)\right)k dk -\frac{L}{2\pi\Xi }.
\end{equation}
Within the strong-coupling virial expansion this term corresponds to the free energy of the system without any ion.

Now we need to compute $U$, remembering that we only need the $L$ dependent terms in $\ln U$. Starting from 
expression (\ref{z1}), we get
\begin{equation}\label{u}
z_1(\xx_0)= 
{Z_0} \exp{(- W(\xx_0))},
\end{equation} 
where $W(\xx_0)$ is an effective one-body potential given by
\begin{equation}\label{u}
W(\xx_0) = - \frac{1}{2}\BB(\xx_0)^T\left(\frac{1}{\alpha}+\AA(\xx_0)\right)^{-1}\BB(\xx_0) + \frac{C'(\xx_0)}{2} + {\textstyle\frac12} {\rm Tr }\log{\left(1+\alpha\AA(\xx_0)\right)}.
\end{equation} 
The $Z_0$ term and the last term in the exponent of the above equation can be rearranged and interpreted in the following way: keeping only terms proportional to $(\nabla \phi)^2$ in the exponential of (\ref{SC_z1}), we can show that
\begin{equation}
\det\left(-\frac{\varepsilon\nabla^2}{4\pi\Xi}\right)^{-1/2}\det\left(1+\alpha\AA(\xx_0)\right)^{-1/2}=\det \left(-\nabla \left[\frac{\varepsilon(\xx)}{4\pi\Xi}+\alpha\delta(\xx-\xx_0)\right]\nabla\right)^{-1/2}.
\label{effdielfghdj}
\end{equation}
This means that these two functional determinants represent the thermal Casimir partition function for a system composed of a finite extension dielectric slab, two semi-infinite dielectric regions outside of it and a single polarizable ion within the slab. The delta function in the expression for the effective dielectric response function on the r.h.s. of (\ref{effdielfghdj}) needs to be regularized to avoid a divergence in the case of a point ion. It is clear that the last term in the effective one-body potential (\ref{u}) describes the thermal Casimir or zero-frequency van der Waals interaction between the polarizable particle and the dielectric interfaces in the system. It is given explicitly by 
\begin{equation}
{\textstyle\frac12} {\rm Tr }\log{(1+\alpha\AA(\xx_0))} = {\textstyle\frac12} {\rm Tr }\log\left(1+\alpha \nabla\nabla'^TG(\xx_0,\xx_0)\right) \simeq {\textstyle\frac12} \alpha {\rm Tr}\left[\nabla\nabla'^TG(\xx_0,\xx_0)\right].
\end{equation}
In the asymptotic regime of large $\xx_0$ we obtain the scaling $\xx_0^{-3}$ which corresponds to the zero-frequency van der Waals interaction between the polarizable particle and a single dielectric discontinuity \cite{Parsegian2005}. Our results are thus completely consistent with everything else we know about the polarizable particles and their zero-frequency van der Waals interactions with dielectric discontinuities \cite{Ninham1997}.

At the end of the calculation we obtain for the $L$-dependent interaction free energy (\ref{fren}) first the usual extensive term, $\frac{L}{2\pi\Xi}$, giving rise to an attractive force between the plates, independent on $L$ while the other terms can not be evaluated analytically but are easily calculated numerically: the computation of $\AA$ follows from (\ref{AA}), $\BB$ is obtained from (\ref{BB}) and finally $C'(\xx_0)$ is obtained in the form
\begin{equation}
C'(\xx_0)=2\Xi\int_0^{k\ind{max}}\frac{\cosh(2kz_0)}{\Delta^{-1}\exp(kL)-\Delta \exp(-kL)}k\, dk + 2L,
\end{equation}
where we differentiated with respect to $L$ before taking the limit $k\rightarrow 0$ in order to get the last term that represents the attraction between the ion and each of the bounding dielectric surfaces of the slab. Finally $U$ is given by integrating (\ref{u}).

\begin{figure*}[t!]
\begin{center}
	\begin{minipage}[b]{0.485\textwidth}
	\begin{center}
		\includegraphics[width=\textwidth]{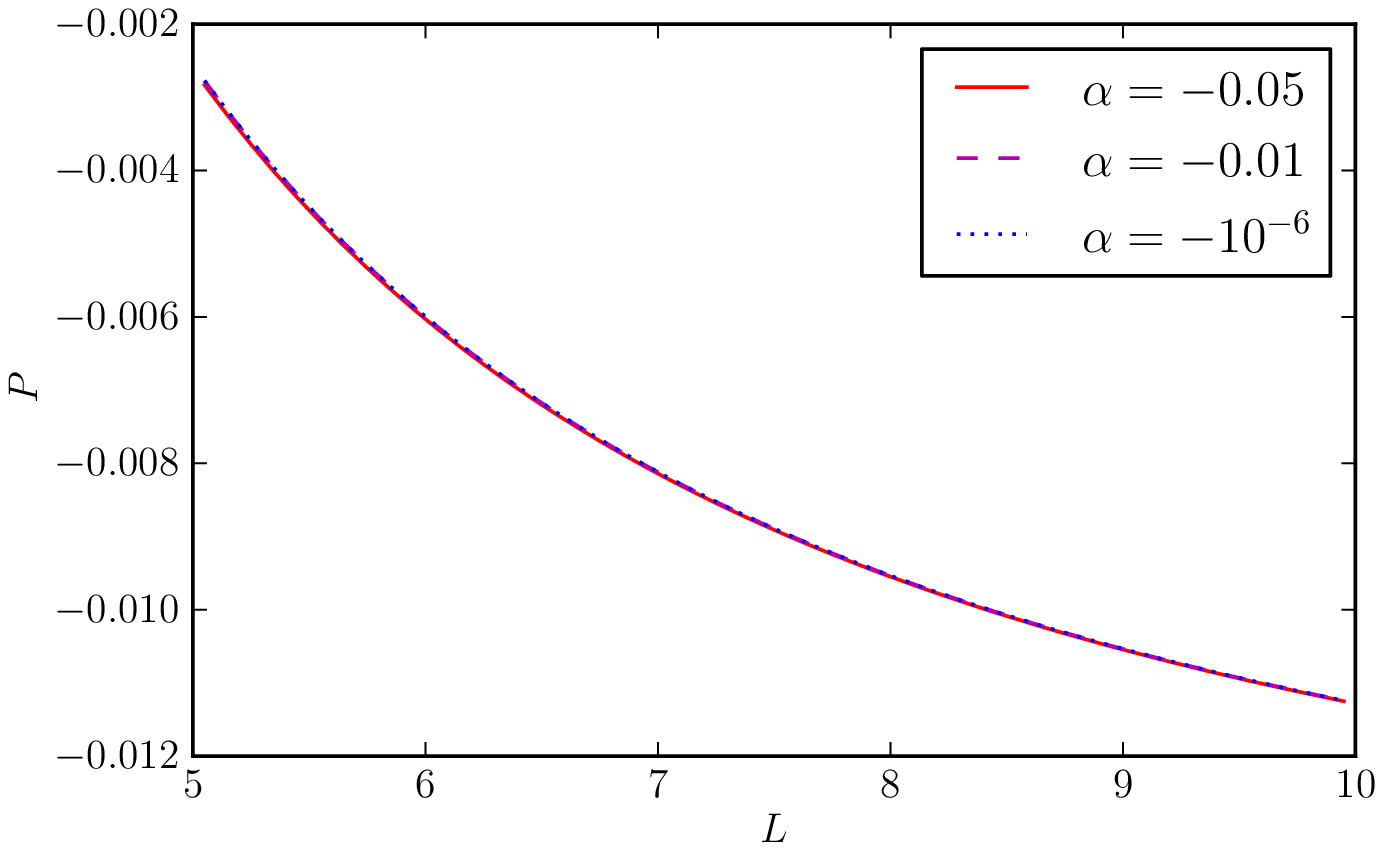}
	\end{center}\end{minipage} \hskip0.25cm
	\begin{minipage}[b]{0.485\textwidth}\begin{center}
		\includegraphics[width=\textwidth]{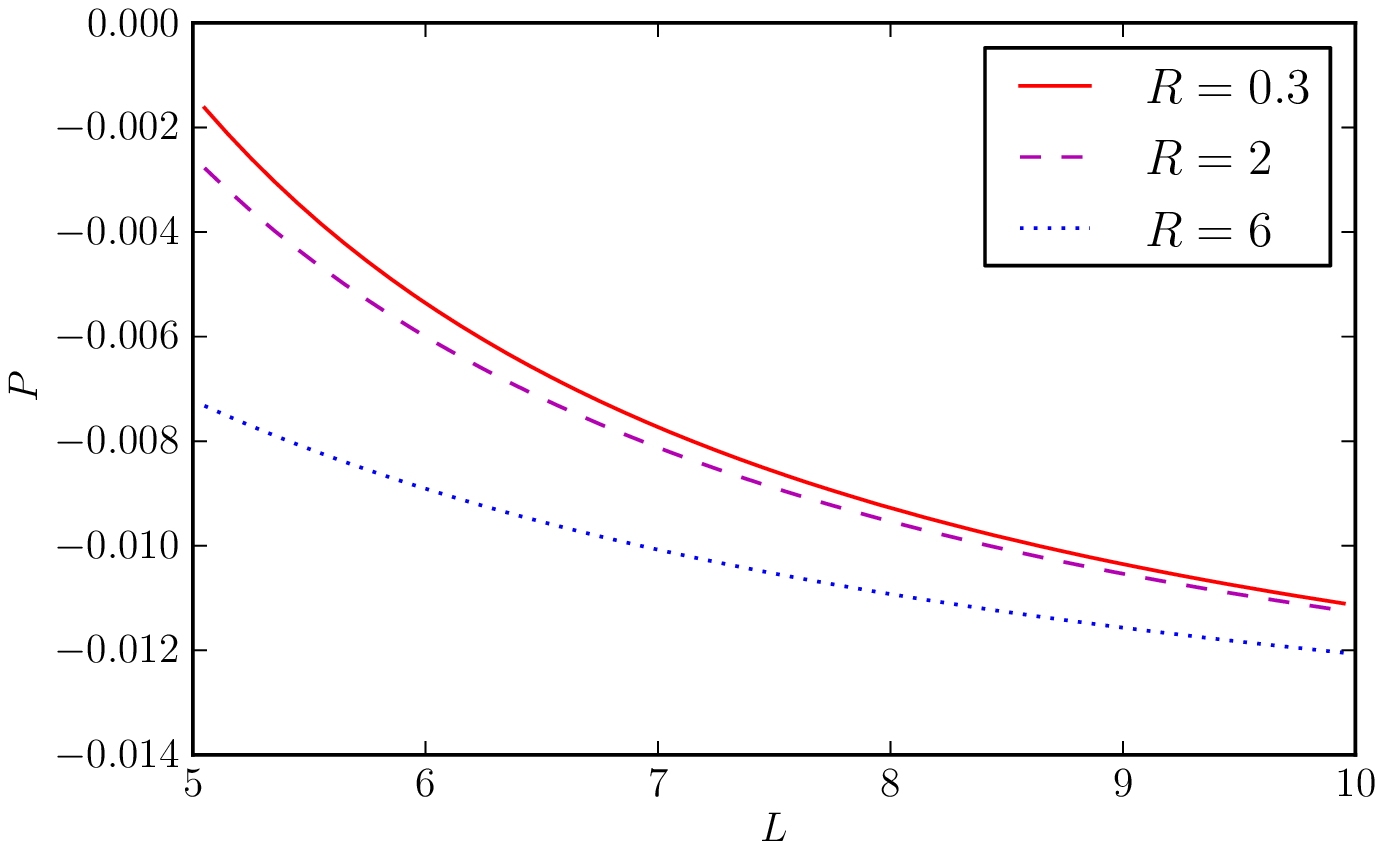}
	\end{center}\end{minipage} \hskip0.25cm	
\caption{Strong coupling pressure as a function of the plate separation $L$. 
\emph{Left}:  $R=2$, $\varepsilon\ind{ext}=0.05$ and $\Xi=10$ for different values of $\alpha$.
\emph{Right}: $\alpha=-0.01$, $\varepsilon\ind{ext}=0.05$ and $\Xi=10$ for different values of $R$.}
\label{f_L_P_sc_alpha_R}
\end{center}\end{figure*}

The pressure obtained from the interaction free energy is shown on Fig. \ref{f_L_P_sc_alpha_R} for different values of $\alpha$ and $R$. The dependance on the polarizability is very weak and non-monotonous, contrarily to what we observed in the weak coupling limit. The results are however very sensitive to the size of the ions, especially for large ions.

\begin{figure*}[t!]
\begin{center}
	\begin{minipage}[b]{0.48\textwidth}
	\begin{center}
		\includegraphics[width=\textwidth]{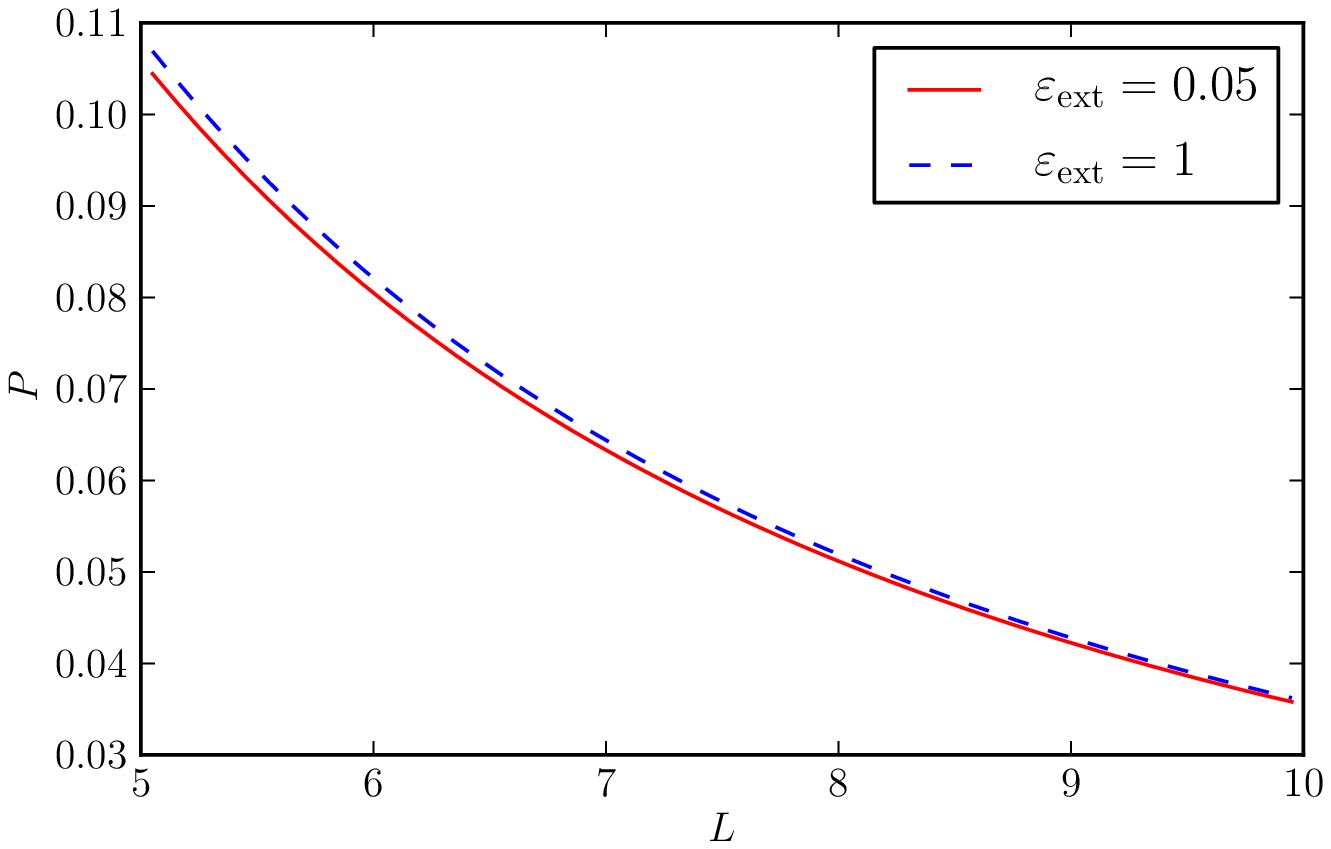}
	\end{center}\end{minipage} \hskip0.25cm
	\begin{minipage}[b]{0.495\textwidth}\begin{center}
		\includegraphics[width=\textwidth]{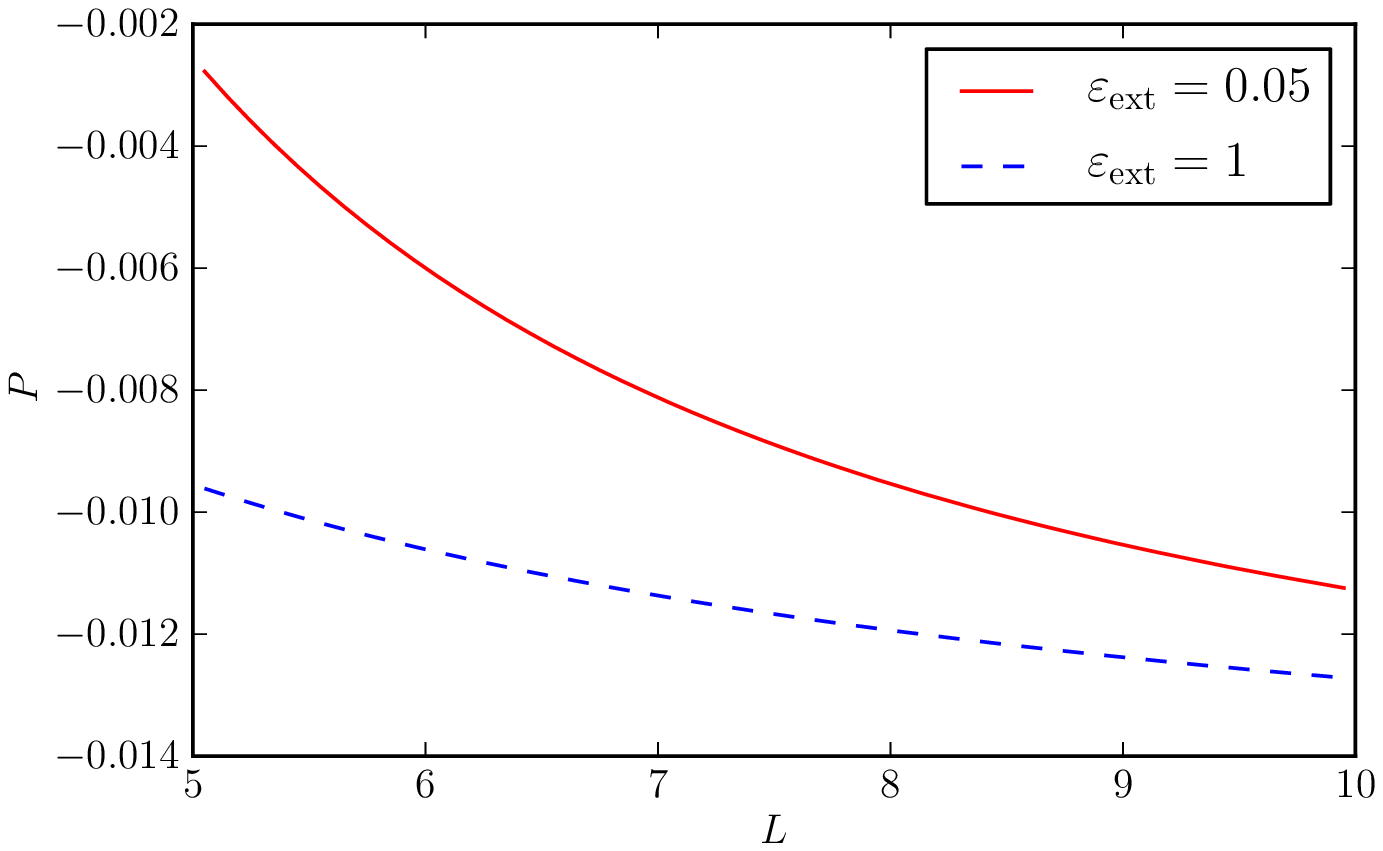}
	\end{center}\end{minipage} \hskip0.25cm	
\caption{Effect of the outer permittivity on the interaction pressure.
\emph{Left}: Weak coupling pressure with $\alpha=-0.1$, $R=1$, $\Xi=0.3$ and $\varepsilon\ind{ext}=0.05$ (solid line) or $\varepsilon\ind{ext}=1$ (dashed line).
\emph{Right}: Strong coupling pressure with $\alpha=-0.01$, $R=2$, $\Xi=10$ and $\varepsilon\ind{ext}=0.05$ (solid line) or $\varepsilon\ind{ext}=1$ (dashed line).
}
\label{f_L_P_eps}
\end{center}\end{figure*}

\section{Discussion and Conclusions}\label{}

In this paper we have formulated a theory of Coulomb fluids that, apart from the charge of the mobile counterions, includes also their static excess polarizability. This leads to a possibility of ion specific effects even for ions with nominally equal valency \cite{Ben-Yaakov2011}. Instead of starting from the phenomenological description of the ionic effects on the local dielectric function, an endeavor pursued in Ref. \cite{Ben-Yaakov2011,Ben-Yaakov2011b}, we rather implemented the effect of ionic polarizability at the level of the field action deriving the appropriate field-theoretic representation of the model. Though this variation in the approach results in the same form of the model at the mean-field level, the formulation presented is eventually more general and suitable for further analysis and implementation of the weak and the strong-coupling asymptotic limits. 

After formulating the model and casting it into a field-theoretic form, we derive the pressure and the ionic density in the mean-field level approximation -- corresponding to the saddle-point of the field-theoretic action. We then add the effects of Gaussian fluctuations of the local electrostatic potential around the mean field. This constitutes the weak-coupling approximation of the complete field theory. 

The effect of the fluctuations around the mean-field saddle point is found to be rather small. In the pressure itself it is barely discernible, see Fig. \ref{f_L_P}, but it does become stronger as the polarizability of the ions is increased, while the density profile shows an effect only very close to the boundaries of the system where the ionic density is enhanced, see Fig. \ref{f_z_rho}, depending on the coupling parameter. This modification of the ionic density in the region close to the dielectric boundaries of the system is partly due to the image effects \cite{Kanduc2007,Jho2008} and partly due to the ionic polarizability. 

We then formulated a full strong-coupling theory which formally corresponds to a single-particle level description and derived its consequences in detail. In the strong-coupling limit the ions are expelled from the vicinity of the dielectric boundaries, see Fig. \ref{f_z_rho}. The origin of this effect lies in the dielectric image interactions that lead to a vicinal exclusion of the ions close to the dielectric discontinuities \cite{Kanduc2007,Jho2008}. 

The results derived here, both for the mean-field plus fluctuations and for the strong coupling regime, exhibit a dependence on the ion polarizability as well as on the size of the ions, Fig. \ref{f_L_P} and \ref{f_L_P_sc_alpha_R}. The effect of ionic polarizability on the interaction pressure is connected partly with the changes in the density profile leading to the changes in the osmotic van't Hoff component of the interactions pressure, and partly with their contribution to the Maxwell stress term in Eq. \ref{presform}. The ionic size dependence comes from divergences naturally present for point dipoles. We have to stress here that what we refer to as the {\em size of the ions} is actually the size of the ionic cavity in the solvent which includes also their hydration shell \cite{Ben-Yaakov2011}. Despite the fact that the field theory arising from our model is a priori independent of the ionic size, we see that if one leaves the domain of mean field theory, by either taking into account fluctuations or going to the strong coupling limit, calculated thermodynamic quantities exhibit ultra-violet, or short distance, divergences. These  divergences are associated with the inclusion of ionic polarizability as they do not arise in strong coupling or in the mean field fluctuations for non-polarizable ions. We have argued that the length scale used to cut-off ultra-violet divergences is thus the {\em size} of the polarizable molecules. Our results are thus in line with Bikerman \cite{Bikerman1942} who long ago argued for the role of the ionic size. The effects of the ion size on the interaction pressure are shown in Fig. \ref{f_L_P_sc_alpha_R}. The size dependence mediated by the polarizability of the ions has nothing to do with steric effects and has not been seen before for non polarizable ions or for polarizable ions on the mean-field level.

The effect of dielectric images, \emph{i.e.} of the outer permittivity, is shown in the case of density on Fig. \ref{f_z_rho_eps}\ and in the case of pressure on Fig. \ref{f_L_P_eps}. The weak coupling limit is only weakly affected by the images, which is to be expected since this regime is dominated by the mean field  that does not depend on $\varepsilon\ind{ext}$ \cite{Kanduc2007}. On the other hand, the strong coupling limit is strongly affected, this time because images add a non negligible term to the correlator (\ref{correl_k}).

Due to the polarizability of the ions, it is also clear that our two approximations break down if the parameters are too extreme, but for different reasons. This is easy to analyze for the strong coupling result (\ref{z1}): here extreme parameter values correspond either to ions which are too small or have too high a (negative) polarizability. 
In this case, the effective permittivity around the ion may turn negative, leading to a field instability that shows up in the partition function.
For the weak coupling limit, on the other hand, if the dielectric function $1+\alpha n(z)$ becomes negative on the mean field level, a divergence appears for the fluctuations about the mean field and the system becomes unstable. 
This can be interpreted in the sense that the validity of the strong coupling vs. weak coupling description no longer depends  on a single coupling parameter, but actually on three parameters. More work would thus be needed to explore different regions of the parameters space and assess the validity of the WC-SC dichotomy in each of them. Our present work can only be seen as a first step towards this complicated endeavor.


One general conclusion stemming from the present work is that the contribution of polarizable counterions to the total partition function is in general non-additive, contrary to what is sometimes assumed \cite{Ninham1997,Edwards2004,LoNostro}. It is in fact highly non-additive at the weak coupling level, whereas it can sometimes be reduced to an additive contribution in the free energy at the strong coupling level, only if the polarizability is large enough. Simply adding a van der Waals ion-polarizability dependent contribution to the electrostatic potential of mean force is simply wrong.

A final  note is in order about the possible computational verifications of our analytical calculations \emph{via} coarse grained simulations that we did not attempt to see through in this work. As polarizability belongs to non-pairwise additive effects the simulation of the present model presents a considerable challenge. One would need to include the image interactions as well as the polarizability couplings to all orders which would appear to be no small accomplishment. Until such time when these type of simulations are actually performed our analytical calculations will remain the sole means to assess the consequences of our model of Coulomb fluids.

\section{Acknowledgments}

VD and DSD would like to thank R.R. Horgan for a discussion about the divergences appearing in the fluctuations about the mean-field. DSD acknowledges support from the Institut Universitaire de France. RP acknowledges support of the The Leverhulme Trust and of ARRS through research program P1-0055 and research project J1-0908.

\bibliographystyle{h-physrev}

\bibliography{biblio3.bib}

\appendix

\section{Fluctuations partition function}\label{ap1}

In this appendix we will compute the path-integral appearing in (\ref{pi_prod}),
\begin{equation}\label{pi}
\mZ^{(2)}_\kk= \int \exp\left(- \frac{S^{(2)}_\kk[\theta]}{\Xi}	\right)[d\theta].
\end{equation}
We introduce the propagator
\begin{equation}
K(\theta_0,\theta_1;z,z';\kk)=\int_{\theta(z)=\theta_0}^{\theta(z')=\theta_1}\exp\left(- \frac{S^{(2)}_{\kk,\textrm{b}}[\theta]}{\Xi}\right)[d\theta],
\end{equation}
where, implicitly, the action is taken only over $[z,z']$. 

We denote by $K\ind{ext}(\theta_0,\theta_1;l;\kk)$ the propagator for a mode $\kk$ on a length $l$ in the external medium; it is given by \cite{Dean2010a}
\begin{equation}
K\ind{ext}(\theta_0,\theta_1;l;\kk)=\sqrt{\frac{\deltext k}{8\pi^2\Xi\sinh(kl)}}\exp\left(-\frac{\deltext k}{8\pi\Xi\tanh(kl)}(\theta_0^2+\theta_1^2)+\frac{\deltext k}{4\pi\Xi\sinh(kl)}\theta_0\theta_1\right).
\end{equation}

We also denote $K_\psi(\theta_0,\theta_1;z,z';\kk)$ the propagator between $z$ and $z'$ in the ionic solution with the field $\psi$ (and the ionic density $n$ contained implicitly in the field). The path-integral (\ref{pi}) is thus 
\begin{eqnarray}\label{xi2_k}
\mZ^{(2)}_\kk & = & \lim_{l\rightarrow\infty}\int K\ind{ext}(\theta_0,\theta_1;l;\kk)K_{\psi\ind{MF}}(\theta_1,\theta_2;0,L;\kk)K\ind{ext}(\theta_2,\theta_3;l;\kk) \exp \left(-\frac{C}{2\Xi} \left[\theta_1^2+\theta_2^2\right]\right)\prod_{j=0}^3d \theta_j,\nonumber\\
& = & \lim_{l\rightarrow\infty}\langle 1|K\ind{ext}(l;\kk)\exp \left(-\frac{C\theta^2}{2\Xi}\right)K_{\psi\ind{MF}}(0,L;\kk)\exp \left(-\frac{C\theta^2}{2\Xi}\right)K\ind{ext}(l;\kk)|1 \rangle,
\end{eqnarray}
where we introduce a matrix notation for the propagator. We have to integrate over the possible outer values of the field, so we need
\begin{equation}\label{kext_s}
\int K\ind{ext}(\theta_0,\theta_1;l;\kk)d\theta_0=\frac{1}{\sqrt{\cosh(kl)}}\exp\left(-\frac{\deltext k\tanh(kl)}{8\pi\Xi}\theta_1^2\right).
\end{equation}
We can see immediately that the limit $l\rightarrow\infty$ is not well defined here: we find $\mZ^{(2)}_\kk=0$. Since we need only the $L$ dependence of $\mZ^{(2)}_\kk$ up to a multiplicative term, we can remove the term $1/\sqrt{\cosh(kl)}$ in the equation above. Thus the path-integral becomes
\begin{eqnarray}\label{Z2_transfer}
\mZ^{(2)}_\kk & = & \lim_{l\rightarrow\infty}\left[\cosh(kl)\left\langle 1\left|K\ind{ext}(l;\kk)\exp \left(-\frac{C\theta^2}{2\Xi}\right)K_{\psi\ind{MF}}(0,L;\kk)\exp \left(-\frac{C\theta^2}{2\Xi}\right)	K\ind{ext}(l;\kk)\right|1 \right\rangle\right] \\
& = & \langle \textrm{ext},\kk|K_{\psi\ind{MF}}(0,L;\kk)|\textrm{ext},\kk\rangle,
\end{eqnarray}
where we have used 
\begin{equation}\label{vect_ext}
|\textrm{ext},\kk\rangle = \exp\left(-\frac{C+\deltext k/4\pi}{2\Xi}\theta^2\right).
\end{equation}

To evaluate the propagator between the plates, we write the bulk action
\begin{equation}
\frac{ S^{(2)}_{\kk,\textrm{b}}[\theta]}{\Xi}=\frac{1}{2}\int_0^L \left[A(z)\theta'(z)^2+B_\kk(z)\theta(z)^2\right]dz,
\end{equation}
with
\begin{eqnarray}
A & = & \frac{1}{2\pi\Xi}\left(\frac{1}{2}+\alpha n+ \alpha^2 n\psi_\textrm{MF}'^2\right), \\
B_\kk & = & \frac{1}{2\pi\Xi}\left[-\frac{\psi_\textrm{MF}''}{2} + \left(\frac{1}{2}+\alpha n\right) k^2\right].
\end{eqnarray}
Then the propagator is of the form \cite{Dean2010a}
\begin{equation}\label{pvv}
K(\theta_0,\theta_1;z,z';\kk)=\sqrt{\frac{b^\kk(z,z')}{2\pi}}\exp\left(-\frac{a\ind{i}^\kk(z,z')}{2}\theta_0^2-\frac{a\ind{f}^\kk(z,z')}{2}\theta_1^2+b^\kk(z,z')\theta_0\theta_1\right).
\end{equation}
It is easy to show that $a\ind{i}^\kk$, $a\ind{f}^\kk$, and $b^\kk$ obey the following composition rules:
\begin{eqnarray}
a^\kk\ind{i}(z,z'+\zeta) & = & a\ind{i}^\kk(z,z')-\frac{b^\kk(z,z')^2}{a^\kk\ind{f}(z,z') + a^\kk\ind{i}(z',z'+\zeta)},\label{comprule}\\ 
a^\kk\ind{f}(z,z'+\zeta) & = & a\ind{f}^\kk(z',z'+\zeta)-\frac{b^\kk(z',z'+\zeta)^2}{a^\kk\ind{f}(z,z') + a^\kk\ind{i}(z',z'+\zeta)},\label{comprule2}\\
b^\kk(z,z'+\zeta) & = & \frac{b^\kk(z,z')b^\kk(z',z'+\zeta)}{a^\kk\ind{f}(z,z') + a^\kk\ind{i}(z',z'+\zeta)}.\label{comprule3}
\end{eqnarray}
On the other hand, one can show that on a small interval $[z,z+\zeta]$, where $A$ and $B_\kk$ are almost constant, they are given by
\begin{eqnarray}
a^\kk\ind{i}(z,z+\zeta) & = & \frac{\sqrt{A(z)B_\kk(z)}}{\tanh(\omega_\kk(z)\zeta)},\\ 
a^\kk\ind{f}(z,z+\zeta) & = & \frac{\sqrt{A(z)B_\kk(z)}}{\tanh(\omega_\kk(z)\zeta)},\\
b^\kk(z,z+\zeta) & = & \frac{\sqrt{A(z)B_\kk(z)}}{\sinh(\omega_\kk(z)\zeta)}, \label{smallint}
\end{eqnarray}
where $\omega_\kk(z)=\sqrt{B_\kk(z)/A(z)}$. From eqs. (\ref{comprule}-\ref{smallint}) we can show that $a^\kk\ind{i}$, $a^\kk\ind{f}$ and $b^\kk$ satisfy
\begin{eqnarray}
\frac{\partial b^\kk}{\partial z'}(z,z') & = & -\frac{a^\kk\ind{f}(z,z')b^\kk(z,z')}{A(z')}, \label{pvv3}\\
\frac{\partial a^\kk\ind{i}}{\partial z'}(z,z') & = & -\frac{b^\kk(z,z')^2}{A(z')}, \label{pvv4}\\
\frac{\partial a^\kk\ind{f}}{\partial z'}(z,z') & = & B_\kk(z') -\frac{a^\kk\ind{f}(z,z')^2}{A(z')}. \label{pvv2}
\end{eqnarray}
with the initial condition
\begin{equation}
b^\kk(z,z')\underset{z'\rightarrow z}{\sim} a^\kk\ind{i}(z,z') \underset{z'\rightarrow z}{\sim} a^\kk\ind{f}(z,z') \underset{z'\rightarrow z}{\sim} \frac{A(z)}{z'-z}
\end{equation}
Eqs. (\ref{pvv}) and (\ref{pvv3}-\ref{pvv2}) are the Pauli-van Vleck formula. We will however use eqs. (\ref{comprule}-\ref{smallint}) for the numerical integration. 

Using (\ref{pvv}) in (\ref{Z2_transfer}), we get
\begin{equation}
\mZ^{(2)}_\kk=\sqrt{\frac{2\pi b^\kk(0,L)}{\left[a\ind{f}^\kk(0,L)+\frac{ C+\deltext k/4\pi}{\Xi}\right]^2-b^\kk(0,L)^2}}.
\end{equation}
However, this expression leads to a pressure that contains a constant term (i.e. independent on $L$), that comes from the fact that in our computation the total volume of the space depends on $L$. This term is thus not physical, and should be removed. Practically, we can notice that it comes from the exponential decay of the function $b^\kk$, $b^\kk(0,L)\sim \exp(-kL)$. The final expression for the path-integral after removing this artificial pressure is thus
\begin{equation}
\mZ^{(2)}_\kk=\exp \left(\frac{kL}{2}\right)\sqrt{\frac{2\pi b^\kk(0,L)}{\left[a\ind{f}^\kk(0,L)+\frac{ C+\deltext k/4\pi}{\Xi}\right]^2-b^\kk(0,L)^2}}.
\end{equation}

\section{Fluctuations correlation function}\label{ap2}

We show here how to compute the $\kk$-mode correlation function $G_\kk(z,z')$ and $\partial\partial'G_\kk(z,z')$ needed in (\ref{rho1+}).

First, we can derive that
\begin{equation}
G_\kk(z,z') = \frac{\langle\textrm{ext},\kk|K_{\psi\ind{MF}}(0,z;\kk)\theta K_{\psi\ind{MF}}(z,z';\kk)\theta K_{\psi\ind{MF}}(z',L;\kk)|\textrm{ext},\kk\rangle}{\langle\textrm{ext},\kk|K_{\psi\ind{MF}}(0,L;\kk)|\textrm{ext},\kk\rangle},
\end{equation}
which can be evaluated explicitly with (\ref{pvv}) and (\ref{vect_ext}) and leads to
\begin{equation}
G_\kk(z,z)=\left(a\ind{f}^\kk(0,z)-\frac{b^\kk(0,z)^2}{\frac{C+\varepsilon\ind{ext}k/4\pi}{\Xi}+a\ind{i}^\kk(0,z)}+[z\rightarrow L-z]\right)^{-1},
\end{equation}
where we took advantage of the symmetry $z\rightarrow L-z$ of the system.

The term $\partial\partial'G_\kk(z,z)$ can then be obtained as the limit
\begin{equation}
\partial\partial'G_\kk(z,z) = \lim_{\zeta\rightarrow 0}(2\zeta)^{-2}[G_\kk(z+\zeta,z+\zeta)-2G_\kk(z-\zeta,z+\zeta)+G_\kk(z-\zeta,z-\zeta)].
\end{equation}
We can again compute this expression analytically, because we have an exact expression for $K_{\psi\ind{MF}}(z,z+\zeta;\kk)$ when $\zeta$ is small. As expected, it diverges. Again, we have to take the size $R$ of the ions into account. To be consistent with the cut-off introduced for pressure, we will regularize the divergence in Fourier space. The point is that locally, the correlator can be computed analytically, and that this local form contains the divergence; thus we will be able to cut it off in the analytic expression.

In this purpose, we decompose the correlator in the following way
\begin{equation}
G_\kk=G_\kk^\mathrm{num}-G_\kk^\mathrm{loc}+G_\kk^\mathrm{loc},
\end{equation}
where $G_\kk^\mathrm{num}$ is the correlator computed using the numerical values for the Pauli-van Vleck functions (dealing with it in Fourier space would need a numerical Fourier transform) and $G_\kk^\mathrm{loc}$ is the "local" correlator, computed assuming that $A(z)$ and $\omega_\kk(z)$ are constant around $z$. Thus the term $G_\kk^\mathrm{num}-G_\kk^\mathrm{loc}$ will not lead to divergences when we will compute its second derivative, and we will be able to write the term $G_\kk^\mathrm{loc}$ in Fourier space and cut its divergence off easily. Explicitly,
\begin{equation}
G_\kk^\mathrm{loc}(z,z')=\frac{\exp(-\omega_\kk(z)|z-z'|)}{2A(z)\omega_\kk(z)},
\end{equation}
its expression in Fourier space is 
\begin{equation}
\tilde G_\kk^\mathrm{loc}(q)=\frac{1}{A(z)(q^2+\omega_\kk(z)^2)},
\end{equation}
so its contribution to the second derivative will be
\begin{eqnarray}
\partial\partial'G_\kk^\mathrm{cut}(z,z) & = & \int_{|q|<q\ind{max}(\kk)} \frac{q^2}{A(z)(q^2+\omega_\kk(z)^2)}\frac{dq}{2\pi} \\ & = & \frac{1}{\pi A(z)}\left[q\ind{max}(\kk)-\omega_\kk(z)\arctan\left(\frac{q\ind{max}(\kk)}{\omega_\kk(z)}\right)\right],
\end{eqnarray}
where the superscript "cut" means that the cut-off has been applied, and the one-dimensionnal cut-off depends on the mode since the three-dimensionnal wave-vector is constrained:
\begin{equation}\label{defqmax}
q\ind{max}(\kk)^2+k^2=k^2\ind{max}.
\end{equation}

Now, we turn to the evaluation of $\partial\partial'G_\kk^\mathrm{num}(z,z)$. The correlators invoked in the above expression can be derived from the quantity
\begin{equation}
Z(a,a',b)=\int \exp\left(-\frac{a}{2}\theta^2-\frac{a'}{2}\theta'^2+b\theta\theta'\right)d\theta d\theta'=\frac{2\pi}{\sqrt{aa'-b^2}},
\end{equation}
where
\begin{eqnarray}
a & = & c+a\ind{i}^\kk(z-\zeta,z+\zeta),\\
a' & = & c'+a\ind{f}^\kk(z-\zeta,z+\zeta),\\
b & = & b^\kk(z-\zeta,z+\zeta),
\end{eqnarray}
with $c=a\ind{f}^\kk(0,z-\zeta)-\frac{b^\kk(0,z-\zeta)^2}{\frac{C+\varepsilon\ind{ext}k/4\pi}{\Xi} +a\ind{i}^\kk(0,z-\zeta)}$ and $c'=a\ind{f}^\kk(0,L-z-\zeta)-\frac{b^\kk(0,L-z-\zeta)^2}{\frac{C+\varepsilon\ind{ext}k/4\pi}{\Xi}+a\ind{i}^\kk(0,L-z-\zeta)}$. We note that the terms appearing in $a$, $a'$ and $b$ can be computed reversing the composition rules (\ref{comprule}-\ref{comprule3}), giving
\begin{eqnarray}
a\ind{i}^\kk(z-\zeta,z+\zeta) & = & \frac{b^\kk(0,z-\zeta)^2}{a\ind{i}^\kk(0,z-\zeta)-a\ind{i}^\kk(0,z+\zeta)}-a\ind{f}^\kk(0,z-\zeta),\\
a\ind{f}^\kk(z-\zeta,z+\zeta) & = & \frac{b^\kk(0,L-z-\zeta)^2}{a\ind{i}^\kk(0,L-z-\zeta)-a\ind{i}^\kk(0,L-z+\zeta)}-a\ind{f}^\kk(0,L-z-\zeta),\\
b^\kk(z-\zeta,z+\zeta) & = & \frac{b^\kk(0,z+\zeta)}{b^\kk(0,z-\zeta)}\left[a\ind{f}^\kk(0,z-\zeta)-a\ind{i}^\kk(z-\zeta,z+\zeta)\right].\\
\end{eqnarray}
where we used $a\ind{f}^\kk(z-\zeta,z+\zeta)=a\ind{i}^\kk(L-z-\zeta,L-z+\zeta)$. Now 
\begin{eqnarray}
G_\kk^\mathrm{num}(z-\zeta,z-\zeta) & = & -2\frac{\partial\ln(Z)}{\partial a} = \frac{a'}{aa'-b^2},\\
G_\kk^\mathrm{num}(z+\zeta,z+\zeta) & = & -2\frac{\partial\ln(Z)}{\partial a'} = \frac{a}{aa'-b^2},\\
G_\kk^\mathrm{num}(z-\zeta,z+\zeta) & = & \frac{\partial\ln(Z)}{\partial b} = \frac{b}{aa'-b^2}.
\end{eqnarray}
We will thus write 
\begin{equation}
[\partial\partial']_\zeta G_\kk^\mathrm{num}(z,z) = \frac{a+a'-2b}{4\zeta^2(aa'-b^2)},
\end{equation}
where $[\partial\partial']_\zeta$ means that the second derivative is computed with a small step $\zeta$. To this quantity, we have to deduce
\begin{equation}
[\partial\partial']_\zeta G_\kk^\mathrm{loc}(z,z) = \frac{1-\exp(-2\omega_\kk(z)\zeta)}{4\zeta^2 A(z)\omega_\kk(z)}.
\end{equation}
Finally, we have
\begin{equation}
\partial \partial'G_\kk(z,z) = [\partial\partial']_\zeta G_\kk^\mathrm{num}(z,z) - [\partial\partial']_\zeta G_\kk^\mathrm{loc}(z,z)  + \partial\partial'G_\kk^\mathrm{cut}(z,z),
\end{equation}
that should not depend on the discretisation step $\zeta$ as soon as it is small enough.

\end{document}